\documentclass[journal]{IEEEtran}

\usepackage[T1]{fontenc}
\usepackage{graphicx}
\usepackage{amsmath,amssymb}
\usepackage{algorithm}
\usepackage{algpseudocode}
\usepackage{booktabs}
\usepackage{array}
\usepackage{multirow}
\usepackage{tabularx}
\usepackage{physics}
\usepackage{cite}
\usepackage{url}
\usepackage{textcomp}
\usepackage{stfloats}
\usepackage{threeparttable}

\begin{document}

    \title{Raw-Curve Quantum Fingerprints: A Mahalanobis Authentication Framework with Drift Early Warning and Adversarial Detection}

\author{Geyuyan Ma,
        Xiangdong Meng,
        Yangyang Fei,
        Zhiqiang Fan,
        Hanshi Zhao,
        Chenhui Wang,
        Haoran Yang,
        Weilong Wang,
        and Zheng Shan*~\thanks{*Corresponding author. Email: shanzhengzz@163.com}%
\thanks{All Authors are with Information Engineering University, Zhengzhou 450001, China.}}

\maketitle

\begin{abstract}
Quantum cloud platforms are poised to deliver powerful computing capabilities, but users have no direct means to verify which physical device executes their workload. This lack of transparency enables hardware substitution attacks, where a malicious adversary could redirect a job to a substituted or inferior processor. We present a general authentication framework that addresses this problem by constructing multi-dimensional quantum fingerprints from raw measurement data. Without any curve fitting, we directly concatenate the raw statistics of complementary experiments into a high-dimensional feature vector that preserves subtle device-specific information. A Mahalanobis nearest-neighbor classifier achieves 100\% benign authentication accuracy on three superconducting processors over a three-week chronological split. The classifier naturally yields an authentication confidence $C_{\mathrm{claimed}}$ which reveals device-specific safety margins and motivates per-device alert thresholds. We assess the framework's robustness under two distinct scenarios. Under additive isotropic Gaussian noise, $C_{\mathrm{claimed}}$ decays predictably at a rate explained by inverse covariance traces, enabling an early warning mechanism. Against white-box adversarial perturbations, the same confidence threshold detects $L_2$ targeted attacks with near-perfect success and reveals device-dependent empirical thresholds for $L_\infty$ attacks, while untargeted and sparse attacks are ineffective. The proposed framework thus unifies fingerprint extraction, drift-resilient authentication, proactive health monitoring, and adversarial defense, offering a practical step toward trustworthy quantum cloud computing.
\end{abstract}


\section{Introduction}
\label{sec:intro}

Quantum computing platforms are rapidly transitioning from laboratory prototypes to cloud-accessible services~\cite{IBMQuantum2024, Preskill2018}. Major cloud providers such as IBM Quantum, Amazon Braket, Google Quantum AI, Microsoft Azure Quantum, and the TianYan quantum cloud platform have made quantum processors available to a growing user base~\cite{Arute2019, Google2019, Amazon2020, Microsoft2021, Tianyan2024}. Meanwhile, a broad spectrum of algorithms and applications---ranging from quantum chemistry simulations~\cite{Peruzzo2014, McClean2016, Georgescu2014} and quantum machine learning~\cite{Biamonte2017, Schuld2015, Bharti2022} to optimization and quantum error correction~\cite{LaRose2020, Knill2005}---are now routinely executed on these platforms. As users submit increasingly sensitive workloads to remote quantum processors, a fundamental security question becomes impossible to ignore: how can users be assured that their computation actually runs on the specific physical device they request, rather than on a substituted or poorly performing one? Quantum processors are fundamentally opaque, so users have no direct visibility into which qubits executed their circuits, nor can they easily detect if their job has been silently redirected to a different device. This lack of transparency creates a critical vulnerability: an adversary could substitute a less capable or intentionally tampered processor, potentially biasing results, leaking information, or reducing the effective computational power. Without reliable hardware authentication, the integrity and trustworthiness of quantum cloud services remain fundamentally compromised. Consequently, practical and robust \emph{quantum hardware authentication} is an essential building block.

One might ask: what physical quantities best serve as stable, unique identifiers for a quantum processor? In principle, parameters such as qubit anharmonicity, cavity resonant frequencies, and individual junction inductances are deeply rooted in fabrication and are highly stable over time. Directly measuring these would yield an ideal hardware fingerprint. However, in a cloud quantum computing environment, users have no low-level access to such internal degrees of freedom. The only interface available is the execution of quantum circuits and the collection of measurement outcomes. Therefore, any practical authentication mechanism must rely solely on the results of programmable quantum experiments. This constraint naturally leads to a design philosophy: instead of chasing a single ``magic'' parameter, we should use a diverse set of quantum experiments to collectively characterize the device from multiple perspectives, forming a multi-dimensional fingerprint that indirectly encodes the underlying hardware features.

In this work we focus on superconducting quantum processors, which are among the most widely deployed platforms for cloud-based quantum computing. Superconducting qubits exhibit device-specific decoherence times, gate fidelities, and frequency detunings that are exquisitely sensitive to fabrication imperfections and local electromagnetic environments~\cite{Krantz2019, Klimov2018}. These physical variations are persistent and practically impossible to clone, making them natural candidates for hardware fingerprints.

Several prior works have explored the use of individual quantum properties for device identification~\cite{Phalak2021, Bathalapalli2023, Arapinis2021}. However, these existing approaches face three critical limitations. First, most of them rely on only one or two types of quantum experiments, thereby failing to capture the full physical complexity of a quantum device. Second, raw measurement curves are typically condensed into a handful of fitted scalar parameters, discarding subtle shape variations that carry additional device-specific information. Third, robustness evaluations are limited to ideal conditions or short time spans (a few days), without quantifying performance degradation under weeks of natural temporal drift, controlled noise, or adversarial input perturbations.

\begin{figure*}[!t]
    \centering
    \includegraphics[width=\textwidth]{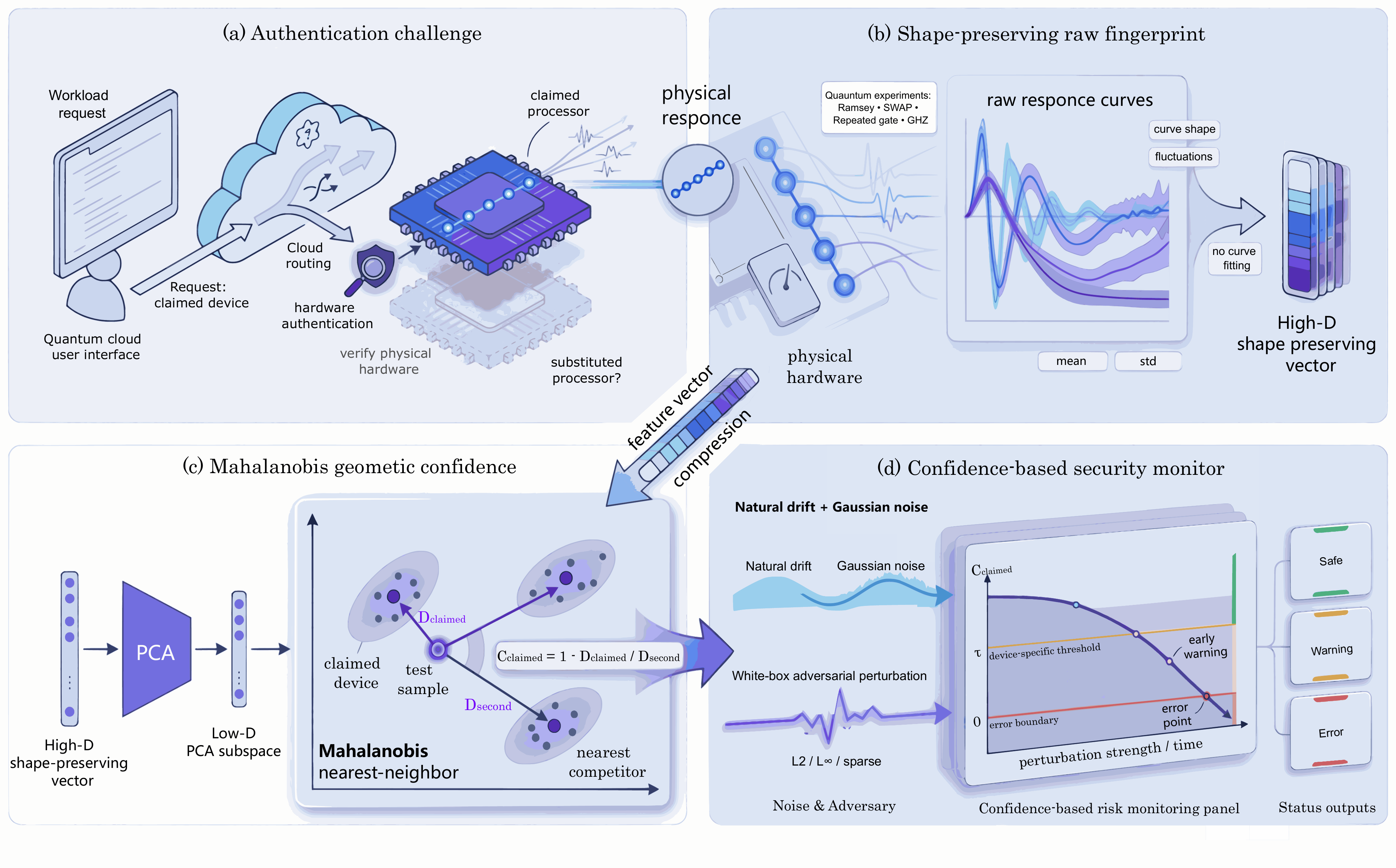}
    \caption{Overview of the proposed quantum hardware authentication framework. (a) A cloud user requests a specific quantum processor; without hardware authentication the job may be redirected to a substituted device. The processor is authenticated by actively probing with four quantum experiments. (b) Raw measurement statistics (means and standard deviations) from the four experiments are directly concatenated into a high-dimensional fingerprint vector without curve fitting, preserving subtle device-specific shape features. (c) PCA compresses the high dimensional fingerprint into a low dimensional subspace; a Mahalanobis nearest-neighbor classifier identifies the claimed device and computes the authentication confidence $C_{\mathrm{claimed}}$ from the distances to the claimed and the nearest competing centroids. (d) The same $C_{\mathrm{claimed}}$ enables unified security monitoring: a device-specific threshold provides early warning under drift and Gaussian noise, and detects white-box adversarial perturbations by rejecting samples with abnormally low confidence.}
    \label{fig:framework}
\end{figure*}

In this paper we present a general framework for quantum hardware authentication built on multi-dimensional raw-curve fingerprints, as shown in Fig.~\ref{fig:framework}. Instead of fitting scalar parameters, we directly concatenate the raw measurement statistics (means and standard deviations) from four complementary quantum experiments into a high-dimensional feature vector, and use a Mahalanobis nearest-neighbor (MNN) classifier for device identification. The four experiments jointly cover single-qubit, two-qubit, and multi-partite entanglement properties. The classifier naturally yields an authentication confidence that quantifies the geometric margin of each decision. This confidence serves three purposes: benign diagnosis, drift monitoring, and adversarial defense. The framework is validated on three distinct superconducting processors over a three-week period with chronological train-test splits, achieving 100\% benign accuracy, predictable confidence decay under additive Gaussian noise that enables early warning, and empirical thresholds against white-box adversarial attacks.

The paper is organized as follows: Section~\ref{sec:dataset} describes the four quantum experiments, the construction of the high-dimensional feature vectors, and analyses justifying the design choices. Section~\ref{sec:mnn} presents the MNN classifier, the geometric separability of the three devices, benign authentication performance, and the distribution of $C_{\mathrm{claimed}}$. Section~\ref{sec:noise} examines the confidence decay under additive isotropic Gaussian noise and validates the early warning mechanism. Section~\ref{sec:adversarial} reports the white-box adversarial evaluation and reveals the device-dependent  empirical thresholds. Section~\ref{sec:conclusion} summarises the contributions and outlines directions for future work.

\section{Feature Extraction and Fingerprint Construction}
\label{sec:dataset}

This section presents the construction of a multi-dimensional raw-curve fingerprint from four complementary quantum experiments---Ramsey interferometry, driven SWAP oscillations, repeated Pauli-X gates, and GHZ entanglement decay, which serves as the foundation of our authentication framework. We first describe the experimental platform and the four experiments, then explain how raw measurement statistics are concatenated into a high-dimensional feature vector without curve fitting. 

\begin{figure*}[!t]
    \centering
    \includegraphics[width=0.85\textwidth]{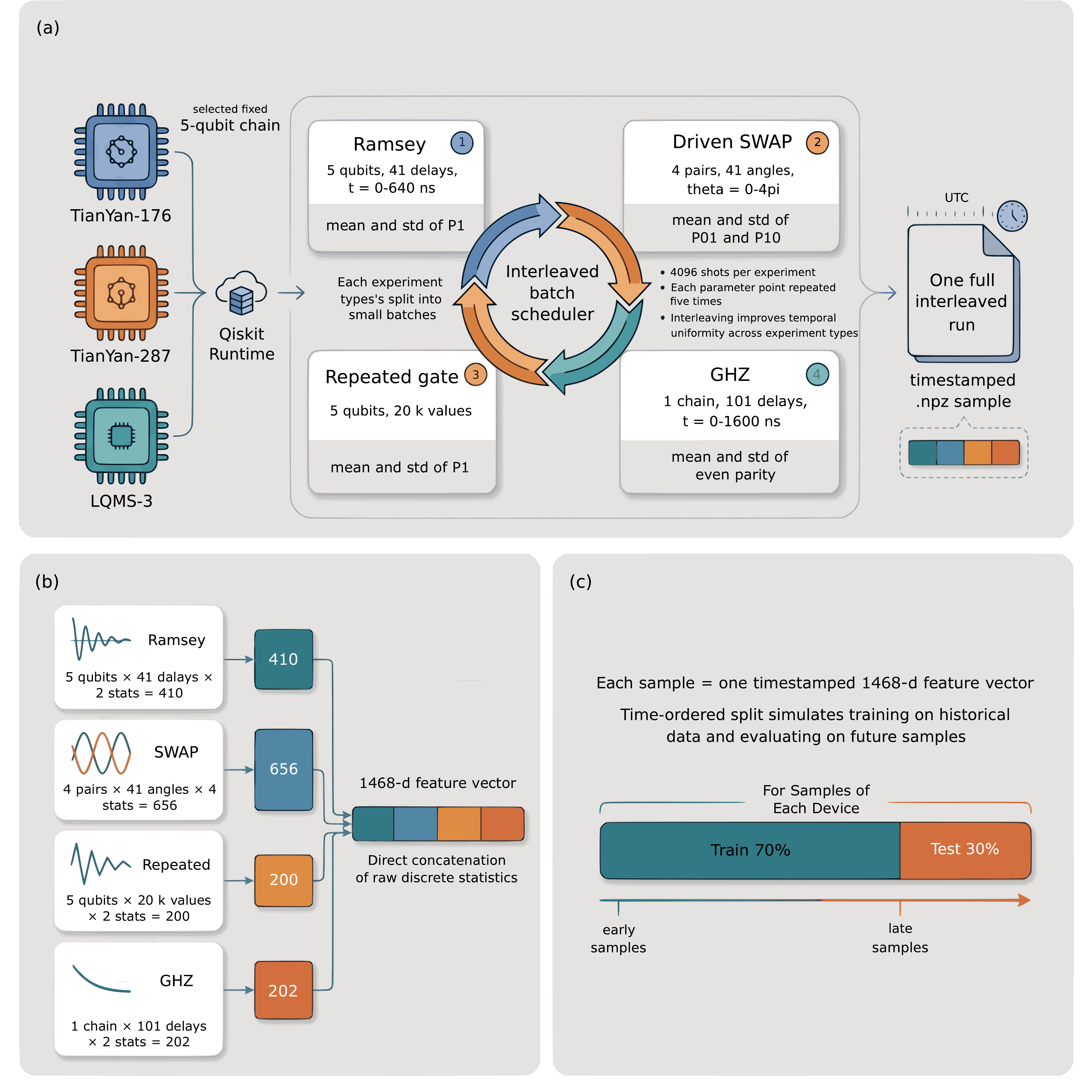}
    \caption{Data acquisition and workflow. (a) Four interleaved quantum experiments are executed on a fixed 5-qubit chain. (b) Raw measurement statistics are directly concatenated into a 1468-dimensional feature vector. (c) The dataset is split chronologically (70\% train / 30\% test).}
    \label{fig:pipeline}
\end{figure*}

\subsection{Experimental Platform and Physical Entity}
\label{sec:device_id}

We validate the proposed framework on three superconducting quantum processors: TianYan-176 and TianYan-287, accessed via the TianYan quantum cloud platform, and LQMS-3, a laboratory-owned processor. On each processor we select a fixed chain of five adjacent qubits as the physical entity to be authenticated. Over a period of approximately three weeks, we collected 40 timestamped samples for TianYan-176, 105 samples for LQMS-3, and 25 samples for TianYan-287. The 5-qubit scale is chosen to keep the experimental overhead manageable for collecting sufficient time-series samples over several weeks. The fingerprinting method is independent of this specific scale; any qubit arrangement capable of supporting the four experiment types can be fingerprinted in the same manner.

\subsection{Interleaved Quantum Experiments}
\label{sec:interleaved}

To capture complementary physical characteristics, we design four experiments covering single-qubit, two-qubit, and multi-partite entanglement properties. The experiments are executed in an interleaved, batch-wise fashion via a scheduler, ensuring that slow environmental drift affects all measurements as uniformly as possible. A complete interleaved run produces a single timestamped \texttt{.npz} file containing all raw measurement curves; this file constitutes one sample in our dataset. Fig.~\ref{fig:pipeline} provides an overview of the acquisition and workflow.

\subsubsection{Ramsey interferometry}
A single qubit is initialised to $|0\rangle$. A $\pi/2$ pulse places it into an equal superposition. After a variable delay $t$ (with a 5\,MHz intentional detuning), a second $\pi/2$ pulse converts the accumulated phase into measurable $|1\rangle$ probability. We sweep $t$ from 0 to 640\,ns in steps of 16\,ns (41 points). All five qubits are characterised independently; for each delay we record the mean and standard deviation of the $|1\rangle$ probability over five repetitions of 4096 shots.

\subsubsection{Driven SWAP oscillations}
A pair of qubits is initialised to $|01\rangle$. The sequence $H\otimes H$, CNOT, $R_z(\theta)$ rotation on the second qubit, CNOT, and $H\otimes H$ implements a tunable SWAP-like interaction. We vary $\theta$ from 0 to $4\pi$ in steps of $\pi/10$ (41 points). For each of the four adjacent qubit pairs, we measure $P_{01}$ and $P_{10}$ and record their mean and standard deviation (five repetitions, 4096 shots). Under ideal coupling these two probabilities are perfectly complementary, but decoherence and readout asymmetry may cause them to deviate, each capturing partially independent fluctuations. We therefore retain both as distinct feature channels.

\subsubsection{Repeated Pauli-X gates}
A single qubit is initialised to $|0\rangle$ and then $k$ consecutive Pauli-X gates are applied. Gate errors cause the outcome probability to degrade nearly toward 0.5 as $k$ increases. We sweep $k$ over 20 selected values: $\{1,2,3,4,5,6,7,8,9,10,12,15,20,25,30,40,50,60,80,100\}$. All five qubits are measured; for each $k$ we record the mean and standard deviation of the $|1\rangle$ probability (five repetitions, 4096 shots).

\subsubsection{GHZ entanglement decay}
A 5-qubit GHZ state $(|00000\rangle+|11111\rangle)/\sqrt{2}$ is prepared, followed by an idle time $t$ (0 to 1600\,ns in steps of 16\,ns, 101 points) during which decoherence degrades the entanglement. Hadamard gates are then applied to all qubits and the even-parity probability is measured. Five repetitions of 4096 shots are performed. The even-parity probability and its standard deviation are retained; the odd-parity values are strictly complementary and carry no additional information.

\begin{table*}[!t]
\centering
\caption{Experiment parameters and extracted statistics.}
\label{tab:exp_params}
\footnotesize
\setlength{\tabcolsep}{5pt}
\renewcommand{\arraystretch}{1.1}
\begin{tabular}{l l l c l}
\toprule
\textbf{Experiment} & \textbf{Qubits/Pairs} & \textbf{Parameter Range} & \textbf{Points} & \textbf{Statistics} \\
\midrule
Ramsey        & 5 qubits & $t \in [0, 640]$\,ns      & 41  & mean, std of $P_1$ \\
SWAP          & 4 pairs  & $\theta \in [0, 4\pi]$ rad & 41  & mean, std of $P_{01}$, $P_{10}$ \\
Repeated Gate & 5 qubits & $k \in \{1,\dots,100\}$    & 20  & mean, std of $P_1$ \\
GHZ           & 5 qubits & $t \in [0, 1600]$\,ns     & 101 & mean, std of even parity \\
\bottomrule
\end{tabular}
\end{table*}

\subsection{Feature Vector Construction}
\label{sec:feature_construction}

From each \texttt{.npz} file we concatenate all raw measurement statistics across the four experiments.

The concatenation yields a 1468-dimensional vector:
\begin{itemize}
    \item Ramsey: 5 qubits $\times$ 41 delays $\times$ 2 statistics = 410.
    \item Driven SWAP: 4 pairs $\times$ 41 angles $\times$ 4 statistics ($P_{01}$ and $P_{10}$ each with mean/std) = 656.
    \item Repeated gate: 5 qubits $\times$ 20 values of $k$ $\times$ 2 statistics = 200.
    \item GHZ: 1 parity channel (even) $\times$ 101 delays $\times$ 2 statistics = 202.
\end{itemize}
Total: $410+656+200+202 = 1468$.

\subsection{Interpretability of the High-Dimensional Features}
\label{sec:interpretability}

The raw 1468-dimensional representation is intentionally rich: neighbouring points on the same decay or oscillation curve are autocorrelated, and some readout channels share common noise sources. This richness ensures that subtle device-specific variations are preserved, even though statistical redundancy and measurement noise are also present. 

To reduce dimensionality and remove statistical correlations, we later compress the vector to 119 dimensions via Principal Component Analysis (PCA) (Section~\ref{sec:pca_mnn}). The geometric interpretability of the fingerprint---specifically, how the four experiment groups contribute to different principal components and how the devices separate in the reduced space---will be analysed in Section~\ref{sec:pca_contrib} and Section~\ref{sec:geometric} using PCA projections and group-contribution analysis.

The robustness advantage of the full fingerprint will be demonstrated under additive isotropic Gaussian noise in Section~\ref{sec:full_vs_mean}.

\section{MNN Authentication and Benign Validation}
\label{sec:mnn}

This section presents the Mahalanobis nearest-neighbor classifier that lies at the heart of our authentication framework, and evaluates its performance under benign (non-adversarial) conditions. We first describe dimensionality reduction and classifier formulation, including the definition of the authentication confidence score $C_{\mathrm{claimed}}$. We then quantify natural temporal drift, analyse the geometric separability of the three devices, and examine the distribution of $C_{\mathrm{claimed}}$ which motivates device-specific alert thresholds.

\subsection{Dimensionality Reduction and Classifier Formulation}
\label{sec:pca_mnn}

The raw 1468-dimensional feature vectors contain statistical correlations and measurement noise. We apply PCA to the standardised training set (earliest 70\% of samples per device). The first 119 principal components retain 95\% of the total variance. The first two components account for 40.1\% and 9.9\% of the variance, respectively (cumulative 50.0\%), and already reveal clear geometric separation among the three devices (Section~\ref{sec:geometric}).

Let the training set consist of samples $\mathbf{x}_i$ with device labels $y_i \in \{c_1, c_2, c_3\}$. After standardisation and PCA projection, each sample is represented as $\mathbf{z}_i \in \mathbb{R}^{d}$ with $d=119$. For each device $c$, we compute the centroid
\[
\boldsymbol{\mu}_c = \frac{1}{|\mathcal{S}_c|}\sum_{i\in\mathcal{S}_c} \mathbf{z}_i,
\]
where $\mathcal{S}_c = \{i : y_i = c\}$. The class-conditional covariance matrix is estimated using the Ledoit--Wolf shrinkage estimator~\cite{Ledoit2004} with a small Tikhonov regularisation ($10^{-6}\mathbf{I}$) to guarantee invertibility (details in Appendix~\ref{app:covariance}). The regularised inverse covariance is denoted $\hat{\mathbf{\Sigma}}_c^{-1}$.

The Mahalanobis distance~\cite{Mahalanobis1936} normalises each direction by its variance and accounts for feature correlations. For $K$ classes with per-class centroids and regularised inverse covariances, the nearest-neighbor rule is the standard quadratic discriminant classifier.

During inference, a test sample $\mathbf{x}$ is standardised and projected to $\mathbf{z}$. Its Mahalanobis distance to each class centroid is
\[
D_M(\mathbf{z}, \boldsymbol{\mu}_c) = \sqrt{(\mathbf{z} - \boldsymbol{\mu}_c)^\top \hat{\mathbf{\Sigma}}_c^{-1} (\mathbf{z} - \boldsymbol{\mu}_c)}.
\]
The rationale for adopting the Mahalanobis distance as the core metric, and its specific roles in classification, confidence evaluation, noise robustness, and adversarial detection, are discussed in Appendix~\ref{app:mahalanobis}.

The sample is assigned to the class with the smallest distance:
\[
\hat{y} = \arg\min_{c} D_M(\mathbf{z}, \boldsymbol{\mu}_c).
\]

The proposed authentication framework operates under a closed-set assumption: all devices that may appear during authentication are known and have been enrolled in the training set. Samples from an unregistered device would be forced into one of the known classes, potentially leading to a false acceptance. Handling open-set authentication (rejecting unknown devices) is left for future work.

\subsection{Quantification of Natural Temporal Drift}
\label{sec:natural_drift}

The data were collected over approximately three weeks, during which the quantum processors underwent routine calibrations and experienced environmental fluctuations. To verify the existence of natural drift and assess its impact on authentication, we compare the Mahalanobis distances of training and test samples to their respective training centroids.

For each device, we split the samples chronologically: the earliest 70\% form the training set and the latest 30\% the test set. Using only the training set, we train the MNN classifier (i.e., obtaining $\boldsymbol{\mu}_c$ and $\hat{\mathbf{\Sigma}}_c^{-1}$ for each device). Then we compute, for every training and test sample belonging to device $c$, its Mahalanobis distance to the training centroid $\boldsymbol{\mu}_c$.

\begin{figure}[!t]
\centering
\includegraphics[width=\columnwidth]{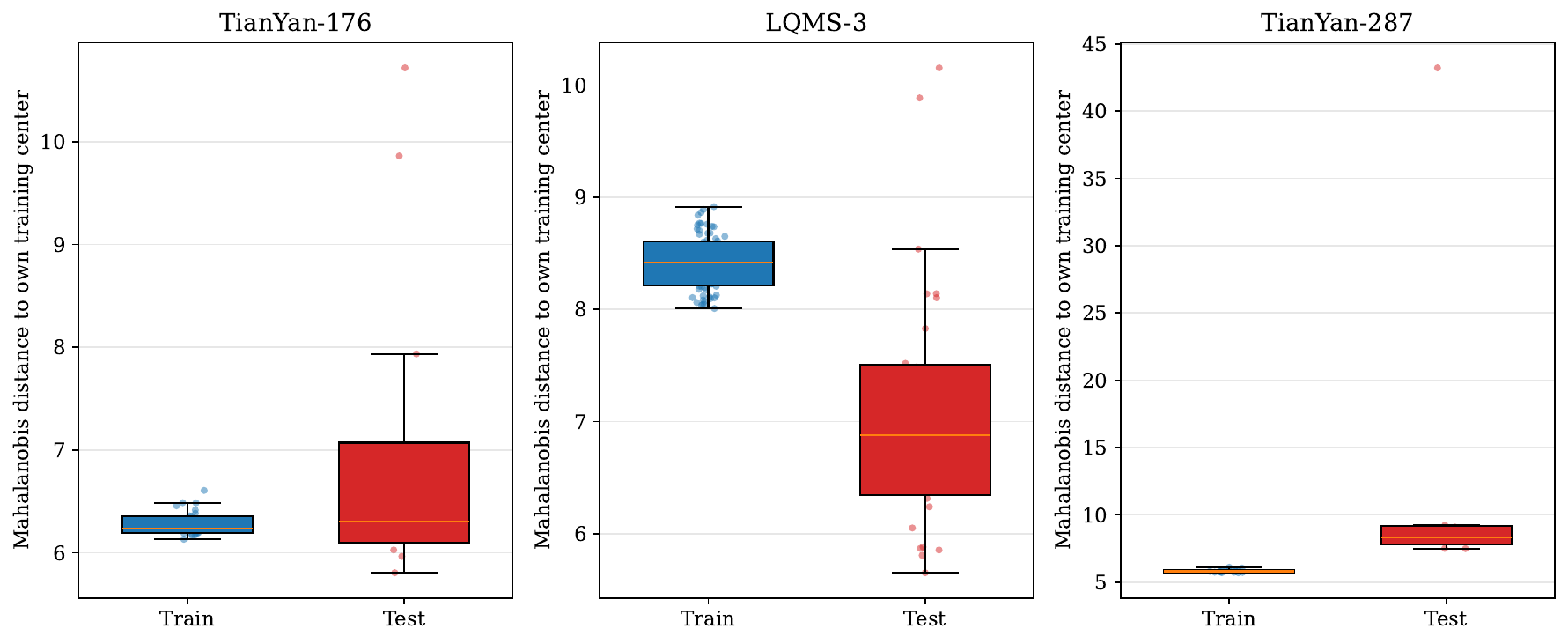}
\caption{Box plots of Mahalanobis distances to the own training centre (blue: training set, red: test set). The central line indicates the median, the box spans the interquartile range (IQR), and the whiskers extend to the farthest data point within $1.5\times\text{IQR}$. Clear differences between training and test distances confirm natural temporal drift on LQMS-3 and TianYan-287; TianYan-176 exhibits a milder shift.}
\label{fig:drift_boxplot}
\end{figure}

Fig.~\ref{fig:drift_boxplot} shows the distributions, and Table~\ref{tab:drift_stats} summarises the statistics.

\begin{table}[!t]
\centering
\caption{Statistics of Mahalanobis distance to own training centre (mean $\pm$ std, median).}
\label{tab:drift_stats}
\footnotesize
\setlength{\tabcolsep}{6pt}
\renewcommand{\arraystretch}{1.1}
\begin{tabular}{l l c c}
\toprule
\textbf{Device} & \textbf{Set} & \textbf{Mean $\pm$ std} & \textbf{Median} \\
\midrule
\multirow{2}{*}{TianYan-176} 
& Train & $6.28 \pm 0.12$ & 6.24 \\
& Test  & $7.05 \pm 1.55$ & 6.31 \\
\midrule
\multirow{2}{*}{LQMS-3} 
& Train & $8.42 \pm 0.24$ & 8.42 \\
& Test  & $7.06 \pm 1.09$ & 6.88 \\
\midrule
\multirow{2}{*}{TianYan-287} 
& Train & $5.84 \pm 0.12$ & 5.80 \\
& Test  & $13.28 \pm 12.24$ & 8.32 \\
\bottomrule
\end{tabular}
\end{table}

For LQMS-3 the test distances are slightly smaller than its training distances, while for TianYan-287 they are significantly larger (containing one extreme outlier with a Mahalanobis distance exceeding 40), confirming temporal drift. The different directions indicate that drift is a random walk around the centroid rather than a monotonic outward shift, as expected under routine environmental fluctuations.

\subsection{PCA-Based Group Contribution Analysis}
\label{sec:pca_contrib}

Before examining the geometric separability of the three devices, we analyse how the four experiment groups contribute to the principal components of the fingerprint space. This analysis provides a physical interpretation of the reduced space: it shows that the four experiments contribute complementary variance components, confirming that the high-dimensional raw-curve fingerprint captures multi-faceted device information.

We apply PCA to the standardised training set comprising all three devices (chronological split, 70\%). To avoid bias from different group sizes (e.g., SWAP has 656 features while GHZ has only 202), we compute the \emph{average squared loading per feature} within each experiment group, then normalise each principal component row-wise. Formally, for principal component $k$ and group $g$, it can be given by
\[
C_{k,g} = \frac{1}{|g|} \sum_{i\in g} (v_{k,i})^2,
\]
where $v_{k,i}$ is the $i$-th component of the $k$-th eigenvector, and rows are scaled so that $\sum_g C_{k,g}=1$. This enables a fair comparison between groups of unequal dimensionality.

\begin{figure}[!t]
\centering
\includegraphics[width=\columnwidth]{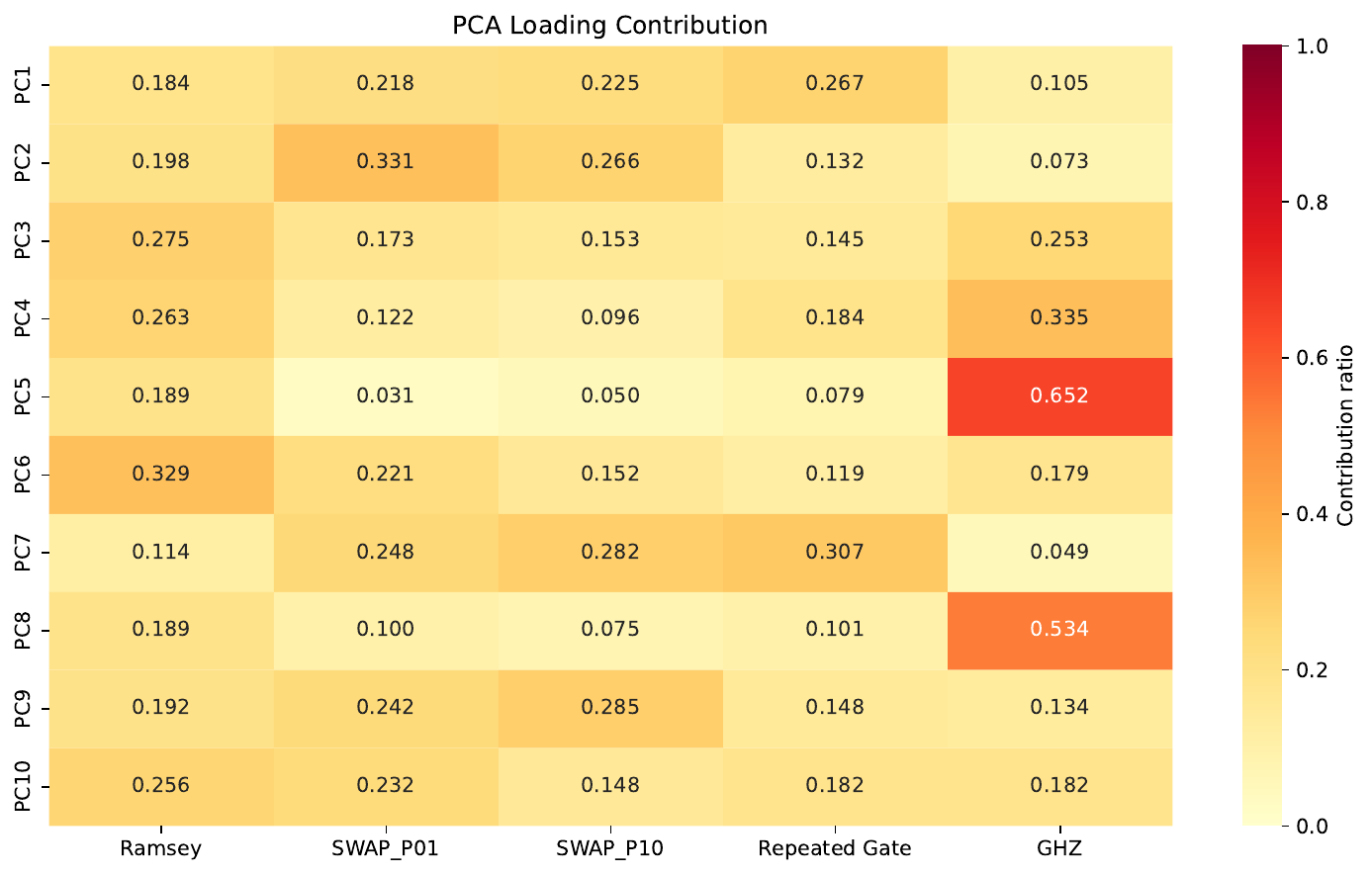}
\caption{Average per-feature contribution of each experiment group to the first ten principal components (row-normalised). Each group contributes to more than one PC, indicating that all four experiments provide distinct variance components.}
\label{fig:pca_heatmap}
\end{figure}

Fig.~\ref{fig:pca_heatmap} shows the average contributions of the first ten principal components. The SWAP experiment is split into $P_{01}$ and $P_{10}$ because they may respond differently to decoherence and readout asymmetry.

Key observations:
\begin{itemize}
    \item \textbf{PC1} is spread across all groups (the contribution of the repeated gate is slightly larger), reflecting a global response common to all measurements.
    \item \textbf{PC2} is dominated by SWAP (especially $P_{01}$), capturing exchange-interaction dynamics.
    \item \textbf{PC3 and PC4} combine Ramsey and GHZ, indicating that single-qubit decoherence and multi-partite entanglement decay are jointly influenced by common environmental fluctuations.
    \item \textbf{PC5 and PC8} are strongly dominated by GHZ (up to 0.652 in PC5), showing that multi-partite entanglement decay carries information not obtainable from single- or two-qubit experiments alone.
    \item \textbf{PC6} is primarily driven by Ramsey, corresponding to single-qubit dephasing and relaxation.
    \item \textbf{PC7} combines SWAP and the repeated gate, suggesting that both experiments may respond to common control-electronics fluctuations.
    \item Although the repeated gate does not independently dominate any of the first ten PCs, it provides persistent secondary contributions across multiple PCs, reflecting discriminative information from gate-error accumulation.
\end{itemize}

Thus, each experiment group contributes to more than one principal component, providing quantitative evidence that the four experiments probe distinct physical aspects and contribute complementary variance components. This lays the foundation for the geometric separability analysis that follows.

\subsection{Geometric Separability}
\label{sec:geometric}

\begin{figure}[!t]
\centering
\includegraphics[width=\columnwidth]{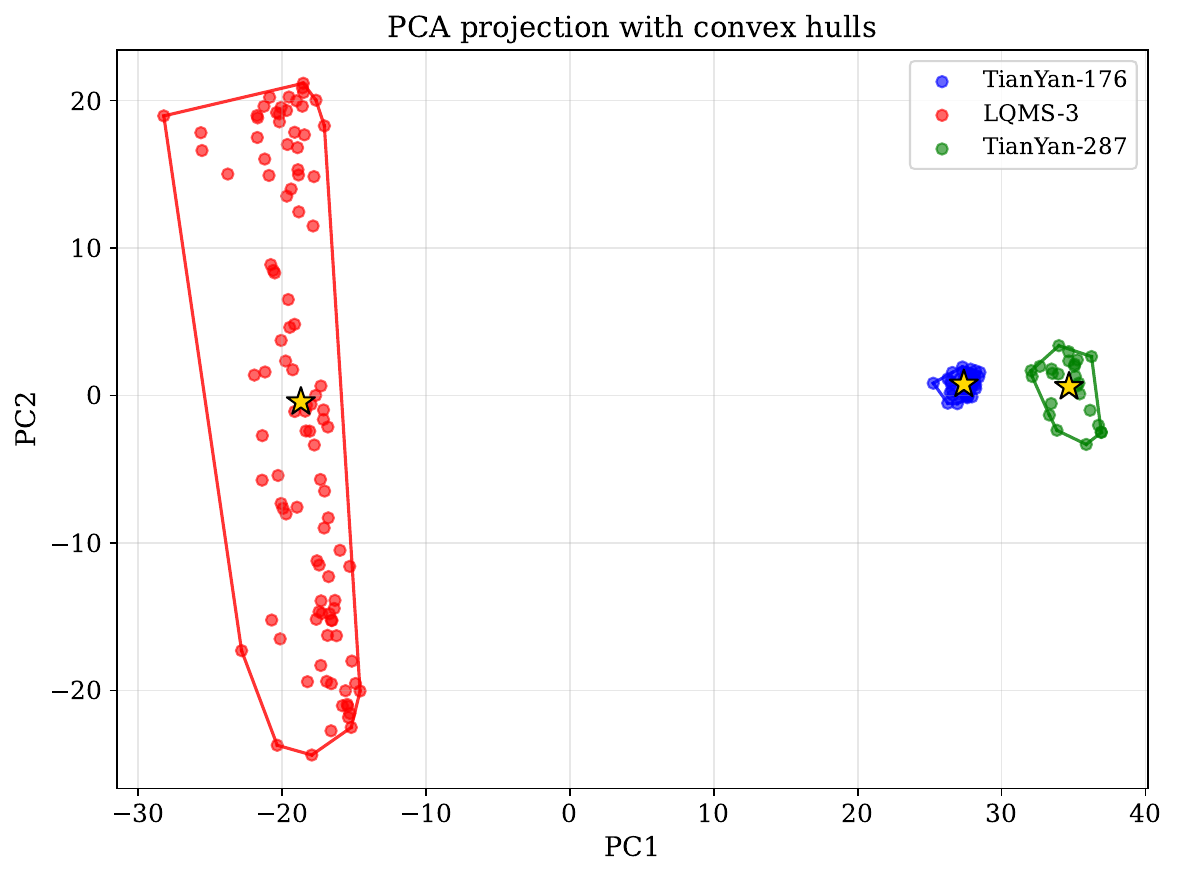}
\caption{PCA projection of all samples onto PC1 and PC2. Solid polygons are convex hulls; star markers denote centroids. LQMS-3 exhibits a wider spread but its centroid is far from the others, yielding high separability.}
\label{fig:pca_convex_hull}
\end{figure}

Fig.~\ref{fig:pca_convex_hull} projects all samples onto PC1 and PC2, together with the convex hull of each device. The apparent proximity of TianYan-176 and TianYan-287 in this two-dimensional view reflects the fact that these two components capture only about half of the total variance (50.0\%); the full 119-dimensional Mahalanobis separability is quantified below.

We compute intra-class distances (sample to own centroid) and inter-class distances (sample to other centroids, averaged over all other devices) in the full 119-dimensional space. Table~\ref{tab:intra_inter} reports the mean $\pm$ standard deviation; Table~\ref{tab:pairwise_ratios} gives pairwise ratios of inter-class distance to the larger intra-class distance. Although LQMS-3 appears to have a very large spread in the PC1-PC2 projection (Fig.~\ref{fig:pca_convex_hull}), its full 119-dimensional intra-class Mahalanobis distance is only moderately larger than that of the other devices (8.42 vs. 6.28 and 5.84). This apparent contradiction is resolved by noting that the Mahalanobis distance normalises each direction by the class‑specific variance. LQMS-3 has a large variance along PC2, so its large scatter in the PCA plot translates into a much smaller contribution to the Mahalanobis distance after division by that variance. The remaining directions do not dominate either; the key factor is the covariance normalization that down‑weights high‑variance directions.

\begin{table}[!t]
\centering
\caption{Intra-class and inter-class Mahalanobis distances (mean $\pm$ std) on training data.}
\label{tab:intra_inter}
\footnotesize
\setlength{\tabcolsep}{6pt}
\renewcommand{\arraystretch}{1.1}
\begin{tabular}{l c c}
\toprule
\textbf{Device} & \textbf{Intra-class distance} & \textbf{Inter-class distance} \\
\midrule
TianYan-176 & $6.28 \pm 0.12$ & $26.86 \pm 11.12$ \\
LQMS-3      & $8.42 \pm 0.24$ & $36.40 \pm 5.12$ \\
TianYan-287 & $5.84 \pm 0.11$ & $35.50 \pm 9.93$ \\
\bottomrule
\end{tabular}
\end{table}

\begin{table}[!t]
\centering
\caption{Pairwise separability: inter-class distance / larger intra-class distance.}
\label{tab:pairwise_ratios}
\scriptsize
\setlength{\tabcolsep}{4pt}
\renewcommand{\arraystretch}{1.1}
\begin{tabular}{l c c c}
\toprule
\textbf{Device pair} & \textbf{Inter-class} & \textbf{Larger intra-} & \textbf{Ratio} \\
 & \textbf{distance} & \textbf{class distance} & \\
\midrule
TianYan-176 vs. LQMS-3    & 39.67 & 8.42 & 4.71 \\
TianYan-176 vs. TianYan-287 & 20.47 & 6.28 & 3.26 \\
LQMS-3 vs. TianYan-287    & 38.61 & 8.42 & 4.58 \\
\bottomrule
\end{tabular}
\end{table}

All pairwise ratios exceed 3.0. A permutation test (10,000 random shuffles of intra- and inter-class distance labels) yields $p < 0.0001$, confirming strong geometric separability. LQMS-3 has the largest intra-class spread yet the highest ratios because its centroid is far from the others. TianYan-287 is more compact but lies closer to TianYan-176 (ratio 3.26), which leads to lower authentication confidence (Section~\ref{sec:confidence}).

\subsection{Benign Authentication Performance}
\label{sec:benign_perf}

\begin{figure}[!t]
\centering
\includegraphics[width=\columnwidth]{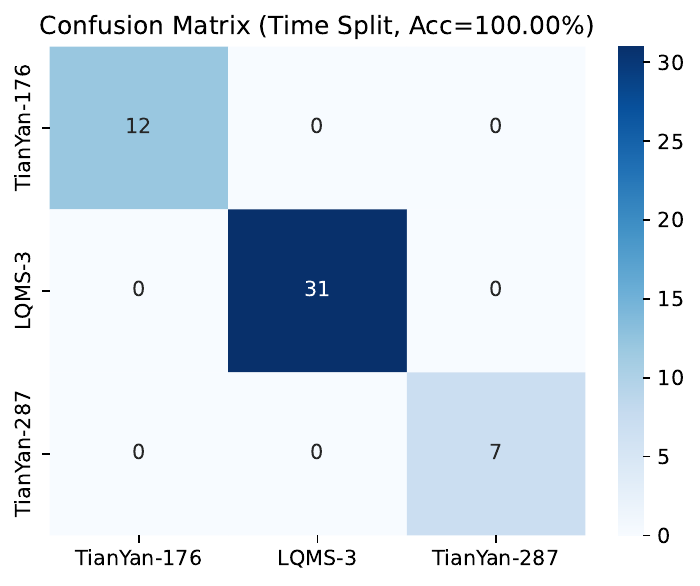}
\caption{Confusion matrix for the chronological test set (50 samples). The colour bar represents the sample count; darker blue indicates a larger number. All samples are correctly classified, with no misclassifications.}
\label{fig:cm}
\end{figure}

We evaluate MNN under three data-split protocols:
\begin{itemize}
    \item \textbf{Chronological split (70\%/30\%):} 100\% accuracy on the 50 test samples. Fig.~\ref{fig:cm} shows the confusion matrix.
    \item \textbf{Random splits (10 independent runs):} Average accuracy $100.00\% \pm 0.00\%$.
    \item \textbf{Leave-one-out cross-validation (LOO-CV):} Accuracies are 100\% for TianYan-176 (40/40), LQMS-3 (105/105), and TianYan-287 (25/25).
\end{itemize}

Given the small number of samples for TianYan-287 (see Section~\ref{sec:device_id}), the LOO-CV estimate necessarily carries a larger variance. Nevertheless, the unanimous agreement across chronological, random, and LOO splits, together with stable and theoretically predictable behaviour under noise and adversarial perturbations reported in later sections, provides consistent evidence that the MNN classifier generalises meaningfully. A definitive assessment would require larger, multi-month datasets, which we identify as an important direction for future work.

\subsection{Authentication Confidence Distribution and Device-Specific Thresholds}
\label{sec:confidence}

In the benign evaluation we possess the true device labels and can therefore compute the authentication confidence that would be observed if a device truthfully claims its identity. For each benign test sample originating from device $c$, we treat $c$ as the claimed identity and compute
\[
C_{\mathrm{claimed}} = 1 - \frac{D_{\mathrm{claimed}}}{D_{\mathrm{second}}},
\]
where $D_{\mathrm{claimed}}$ is the distance to the centroid of device $c$ and $D_{\mathrm{second}}$ is the distance to the nearest centroid among the remaining devices. This value directly quantifies the geometric margin: $C_{\mathrm{claimed}} \approx 1$ indicates a large margin, while $C_{\mathrm{claimed}} \approx 0$ signals proximity to the decision boundary. A negative value means the sample is closer to a wrong centroid---a situation that does not occur in the benign test set.

\begin{figure}[!t]
\centering
\includegraphics[width=\columnwidth]{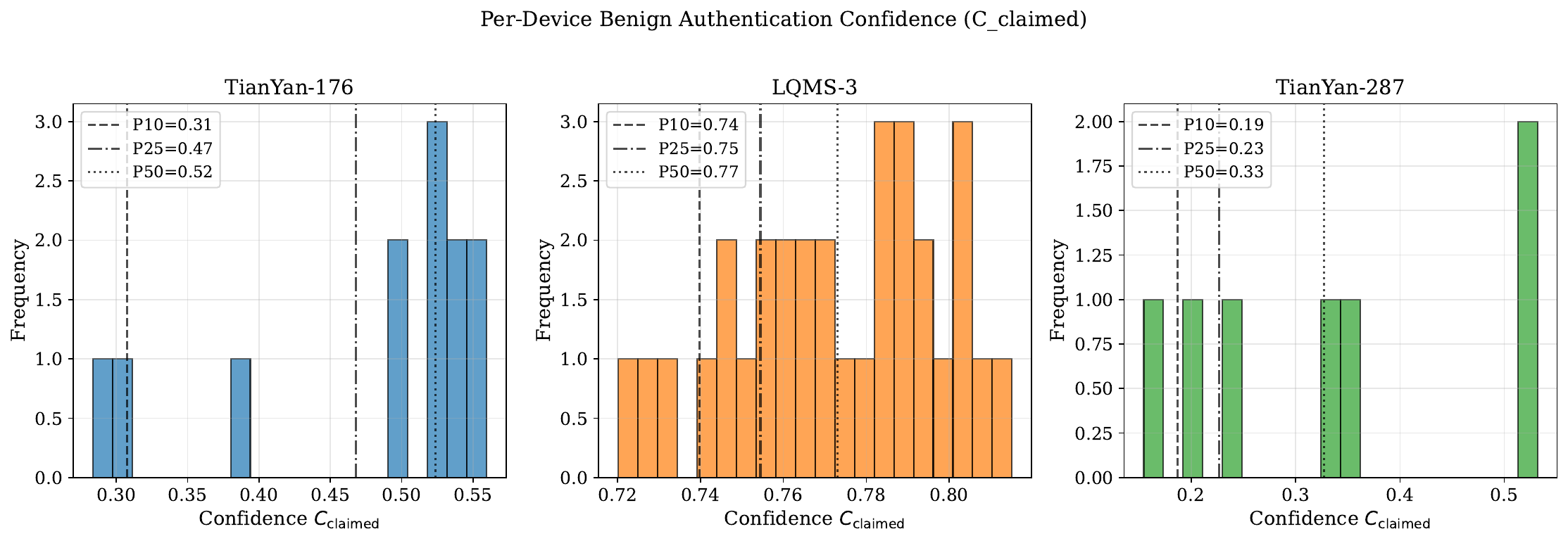}
\caption{Distribution of authentication confidence $C_{\mathrm{claimed}}$ on benign test samples. Vertical lines mark the 10th, 25th and 50th percentiles. LQMS-3 exhibits high confidence (median 0.77), TianYan-176 moderate (0.52), and TianYan-287 low (0.33).}
\label{fig:conf_hist}
\end{figure}

The distribution of $C_{\mathrm{claimed}}$ thus reflects the natural geometric margin of each device. Fig.~\ref{fig:conf_hist} shows histograms with the 10th, 25th and 50th percentiles; Table~\ref{tab:conf_percentiles} lists the numerical values.

\begin{table}[!t]
\centering
\caption{Percentiles of benign authentication confidence $C_{\mathrm{claimed}}$ per device.}
\label{tab:conf_percentiles}
\footnotesize
\setlength{\tabcolsep}{4pt}
\renewcommand{\arraystretch}{1.1}
\begin{tabular}{l c c c c c}
\toprule
\textbf{Device} & \textbf{P1} & \textbf{P5} & \textbf{P10} & \textbf{P25} & \textbf{P50} \\
\midrule
TianYan-176 & 0.285 & 0.292 & 0.308 & 0.468 & 0.524 \\
LQMS-3      & 0.723 & 0.729 & 0.740 & 0.754 & 0.773 \\
TianYan-287 & 0.158 & 0.171 & 0.187 & 0.227 & 0.328 \\
\bottomrule
\end{tabular}
\end{table}

LQMS-3, with the largest geometric margins, has consistently high confidence (median 0.77); TianYan-176 moderate (0.52); TianYan-287, closest to TianYan-176, exhibits the lowest confidence (median 0.33) and a wider spread. These device-specific distributions motivate per-device alert thresholds, employed in Section~\ref{sec:noise} for early warning.

\section{Confidence Decay and Early Warning under Additive Isotropic Gaussian Noise}
\label{sec:noise}

In Section~\ref{sec:natural_drift} we demonstrated that the MNN classifier maintains perfect accuracy under weeks of natural temporal drift. This section examines the behaviour of the authentication confidence $C_{\mathrm{claimed}}$ when test samples are subjected to controlled additive isotropic Gaussian noise. We show that $C_{\mathrm{claimed}}$ decreases predictably as the noise strength increases, and that this decay enables an early warning mechanism through device-specific alert thresholds.

\subsection{Experimental Setup}
\label{sec:iso_setup}

We use the same chronological 70/30 train-test split as before. Isotropic Gaussian noise $\boldsymbol{\eta} \sim \mathcal{N}(\mathbf{0}, \sigma^2 \mathbf{I})$ is added to each test sample. To ensure a uniform interpretation of $\sigma$ across all 1468 feature dimensions, we scale $\sigma$ by the per-feature standard deviation estimated from the training set. Concretely, for each feature dimension $j$, the added noise $\eta_j$ has variance $\sigma^2 \cdot s_j^2$, where $s_j$ is the training-set standard deviation of that feature. Thus $\sigma=1$ corresponds to noise whose typical magnitude equals the natural measurement fluctuation of that feature.

Additive isotropic Gaussian noise serves as a standardized stress test for classifier robustness against independent measurement fluctuations. While it does not capture all complexities of real quantum noise, it is a well-established benchmark for sensitivity analysis. In superconducting qubit systems, noise sources such as readout errors, fast environmental variations are often approximated as Gaussian processes~\cite{Krantz2019, Burnett2019, Klimov2018}. Hence, evaluating our framework under this controlled perturbation provides a meaningful indicator of its resilience to measurement imperfections.

For each $\sigma \in \{0,1,\dots,10\}$, we generate 10 independent noise realisations. The range $\sigma=0$ to $10$ spans from no added noise to levels where authentication accuracy substantially degrades, allowing a complete observation of the confidence decay trajectory.

\subsection{Confidence Decay and Theoretical Explanation}
\label{sec:iso_results}

\begin{figure}[!t]
\centering
\includegraphics[width=\columnwidth]{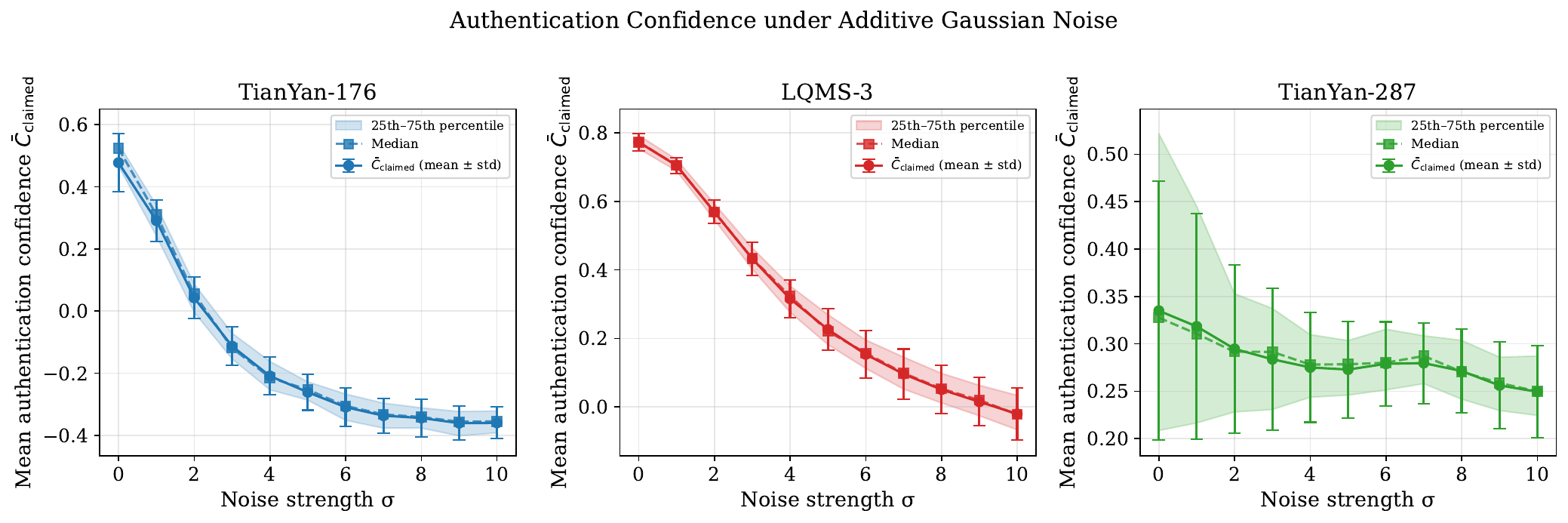}
\caption{Mean authentication confidence $\bar{C}_{\mathrm{claimed}}$ under additive isotropic Gaussian noise (error bars: one standard deviation; shaded region: 25th--75th percentiles).}
\label{fig:c_vs_sigma}
\end{figure}

Fig.~\ref{fig:c_vs_sigma} shows the experimental mean $\bar{C}_{\mathrm{claimed}}$ as a function of $\sigma$. All three devices exhibit an overall decreasing trend, but with markedly different rates: TianYan-176 drops rapidly and some samples become negative near $\sigma=2$; LQMS-3 decreases gradually, staying positive even at very high noise level; TianYan-287 declines extremely slowly and remains positive throughout.

\begin{table}[!t]
\centering
\caption{Mean $\bar{C}_{\mathrm{claimed}}$ and accuracy at selected $\sigma$.}
\label{tab:c_stats}
\footnotesize
\setlength{\tabcolsep}{6pt}
\renewcommand{\arraystretch}{1.1}
\begin{tabular}{l c c c}
\toprule
\textbf{Device} & $\sigma$ & \textbf{Mean $\pm$ std} & \textbf{Accuracy} \\
\midrule
\multirow{3}{*}{TianYan-176}
& 0 & $0.478 \pm 0.094$ & 1.000 \\
& 2 & $0.043 \pm 0.067$ & 0.725 \\
& 5 & $-0.261 \pm 0.058$ & 0.000 \\
\midrule
\multirow{4}{*}{LQMS-3}
& 0 & $0.773 \pm 0.025$ & 1.000 \\
& 2 & $0.569 \pm 0.034$ & 1.000 \\
& 6 & $0.153 \pm 0.069$ & 0.977 \\
& 10& $-0.021 \pm 0.077$ & 0.400 \\
\midrule
\multirow{3}{*}{TianYan-287}
& 0 & $0.335 \pm 0.137$ & 1.000 \\
& 2 & $0.295 \pm 0.089$ & 1.000 \\
& 10& $0.250 \pm 0.049$ & 1.000 \\
\bottomrule
\end{tabular}
\end{table}

For a test sample with noise-free Mahalanobis distance $D_{c'}(0)$ to the centroid of device $c'$, the expected distance under isotropic Gaussian noise of strength $\sigma$ can be expressed as
\[
\widehat{D}_{c'}(\sigma) = \sqrt{ D_{c'}^{2}(0) + \sigma^{2} \operatorname{tr}(\boldsymbol{\Sigma}_{c'}^{-1}) },
\]
where $\operatorname{tr}(\boldsymbol{\Sigma}_{c'}^{-1})$ is the trace of the regularised inverse covariance matrix of device $c'$. The predicted authentication confidence is then
\[
\widehat{C}_{\mathrm{claimed}}(\sigma) = 1 - \frac{ \widehat{D}_{c}(\sigma) }{ \min_{c'\neq c} \widehat{D}_{c'}(\sigma) }.
\]
Using the training-set traces ($\operatorname{tr}(\boldsymbol{\Sigma}_{176}^{-1})=52.11$, $\operatorname{tr}(\boldsymbol{\Sigma}_{287}^{-1})=26.68$, $\operatorname{tr}(\boldsymbol{\Sigma}_{\mathrm{LQMS}}^{-1})=36.88$) and the noise-free distances from Table~\ref{tab:drift_stats}, the formula explains the three distinct behaviours in Fig.~\ref{fig:c_vs_sigma}:

\begin{itemize}
    \item \textbf{TianYan-176}: its large own trace (52.11) and small competitor trace (26.68) cause the ratio $\widehat{D}_{\mathrm{claimed}}/\widehat{D}_{\mathrm{second}}$ to rise quickly, so $\widehat{C}_{\mathrm{claimed}}$ drops sharply, turning negative near $\sigma=2$.
    \item \textbf{LQMS-3}: despite a larger own trace (36.88 vs. 26.68), the huge initial competitor distance keeps the ratio growth very slow, so $\widehat{C}_{\mathrm{claimed}}$ declines gradually, staying positive until extreme noise.
    \item \textbf{TianYan-287}: its own trace (26.68) is smaller than the competitor's (52.11), yet its initial claimed distance is also substantially smaller, yielding a positive monotonicity indicator $\Delta = bc - ad > 0$ (see Appendix~\ref{app:monotonicity}). The ratio increases only slightly, so $\widehat{C}_{\mathrm{claimed}}$ decreases very slowly toward a positive asymptotic limit. The total expected drop is small, matching the nearly flat experimental curve.
\end{itemize}

A complete derivation of the expected distance formula and the monotonicity condition is provided in Appendix~\ref{app:theory}.

\subsection{Early Warning Mechanism}
\label{sec:early_warning}

The predictable decay of $C_{\mathrm{claimed}}$ enables a simple early warning strategy.

We set a device-specific alert threshold $\tau_c$ as the 5th percentile (P5) of the benign $C_{\mathrm{claimed}}$ distribution (Fig.~\ref{fig:conf_hist}, Table~\ref{tab:conf_percentiles}). This choice gives a 5\% false-alarm rate under normal operation. The thresholds are $\tau_{176}=0.292$, $\tau_{\mathrm{LQMS}}=0.729$, $\tau_{287}=0.171$.

For a sample claiming identity $c$:
\begin{itemize}
    \item \textbf{Safe}: $C_{\mathrm{claimed}} \ge \tau_c$ --- no action.
    \item \textbf{Warning}: $0 < C_{\mathrm{claimed}} < \tau_c$ --- still correctly authenticated but suspicious; system raises an alert and may recommend re-calibration (this re-calibration refers to the two-stage procedure mentioned later.).
    \item \textbf{Error}: $C_{\mathrm{claimed}} \le 0$ --- authentication fails.
\end{itemize}
Since correct authentication implies $C_{\mathrm{claimed}} > 0$, Warning always precedes Error under a decreasing confidence trajectory. The warning mechanism does not require monotonic decay; it triggers solely based on the current confidence relative to the device‑specific threshold, making it effective even under fluctuating noise.

\begin{figure}[!t]
\centering
\includegraphics[width=\columnwidth]{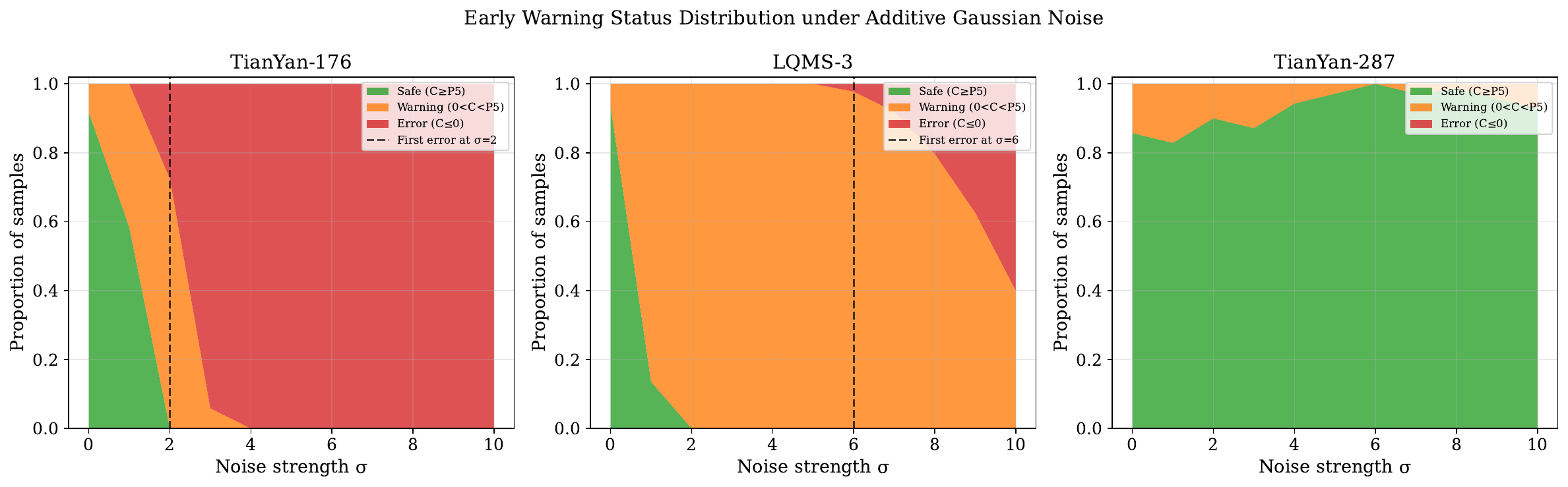}
\caption{Stacked area plot of warning states. Green: Safe; orange: Warning; red: Error. Dashed lines mark the first $\sigma$ with non-zero error.}
\label{fig:early_warning_stack}
\end{figure}

We count the proportion of samples in each state using the noisy test samples from Section~\ref{sec:iso_results}. Fig.~\ref{fig:early_warning_stack} shows the stacked areas; Table~\ref{tab:early_warning} gives representative values (full table in Appendix~\ref{app:ch4_tables}).

\begin{table}[!t]
\centering
\caption{Proportion of samples in each state at selected $\sigma$ (threshold $\tau_c$ = benign P5).}
\label{tab:early_warning}
\footnotesize
\setlength{\tabcolsep}{5pt}
\renewcommand{\arraystretch}{1.1}
\begin{tabular}{l c c c c}
\toprule
\textbf{Device ($\tau_c$)} & $\sigma$ & \textbf{Safe} & \textbf{Warning} & \textbf{Error} \\
\midrule
\multirow{3}{*}{TianYan‑176 (0.292)}
& 0 & 0.917 & 0.083 & 0.000 \\
& 1 & 0.583 & 0.417 & 0.000 \\
& 3 & 0.000 & 0.058 & 0.942 \\
\midrule
\multirow{3}{*}{LQMS‑3 (0.729)}
& 0 & 0.936 & 0.064 & 0.000 \\
& 1 & 0.136 & 0.864 & 0.000 \\
& 5 & 0.000 & 1.000 & 0.000 \\
\midrule
\multirow{3}{*}{TianYan‑287 (0.171)}
& 0 & 0.857 & 0.143 & 0.000 \\
& 1 & 0.829 & 0.171 & 0.000 \\
& 10& 0.929 & 0.071 & 0.000 \\
\bottomrule
\end{tabular}
\end{table}

For TianYan-176, the warning proportion reaches 41.7\% at $\sigma=1$ while all samples remain correctly authenticated; errors first appear at $\sigma=2$. For LQMS-3, 100\% of the samples enter Warning from $\sigma=2$ to $\sigma=5$ yet authentication remains perfect, providing a clean four-$\sigma$ recalibration window. TianYan-287 never enters Error and rarely triggers Warning, as its competitor's much larger trace keeps $C_{\mathrm{claimed}}$ stable.

When a warning is raised, the recommended procedure consists of two stages:
\begin{itemize}
    \item \textbf{Standard qubit calibration.} This corrects control-parameter drifts that may have caused the confidence drop.
    \item \textbf{Verification with fresh samples.} A small set of new samples is collected from the alerted device. If their confidence returns to the Safe region ($C_{\mathrm{claimed}} \ge \tau_c$), the existing enrollment remains valid. Otherwise, a \emph{full re-enrollment} is performed: new samples are collected from \emph{all} enrolled devices, and the entire framework (PCA subspace, centroids, covariances, and alert thresholds) is recomputed.
\end{itemize}

\subsection{Benefit of Raw-Curve Fingerprints under Noise}
\label{sec:full_vs_mean}

Recall that our full fingerprint includes both mean values and per-point standard deviations (Section~\ref{sec:feature_construction}). To assess the contribution of the standard deviation features, we compare the authentication accuracy under the same isotropic Gaussian noise when using only the 734 mean values (i.e., discarding all standard deviations) versus the full fingerprint.

Table~\ref{tab:full_vs_mean} shows the accuracy for noise strengths from $\sigma = 0$ to $5$ (the practically relevant regime where the early warning mechanism is still active). The full fingerprint significantly outperforms the mean‑only fingerprint in the moderate noise regime ($\sigma=2$–$4$). At $\sigma=2$, the full fingerprint maintains \textbf{100\%} accuracy while the mean‑only fingerprint drops to 80.4\%. Even at $\sigma=3$, the full fingerprint still achieves \textbf{96.6\%}, compared to only 76.0\% for the mean‑only version.

This demonstrates that retaining the raw measurement fluctuations (standard deviations) substantially improves robustness against Gaussian noise, justifying the design of a high‑dimensional raw‑curve fingerprint. The full fingerprint provides a critical safety margin when the device operates in noisy environments.

\begin{table}[!t]
\centering
\caption{Authentication accuracy under isotropic Gaussian noise: full fingerprint (1468‑dim) vs only mean values (734‑dim).}
\label{tab:full_vs_mean}
\footnotesize
\setlength{\tabcolsep}{5pt}
\begin{tabular}{c c c}
\toprule
$\sigma$ & Full fingerprint & Only mean \\
\midrule
0 & 1.000 & 1.000 \\
1 & 1.000 & 1.000 \\
2 & \textbf{1.000} & 0.804 \\
3 & \textbf{0.966} & 0.760 \\
4 & 0.846 & 0.760 \\
5 & 0.772 & 0.760 \\
\bottomrule
\end{tabular}
\end{table}

\subsection{Comparison with a Single-Class Baseline}
\label{sec:comp_baseline}

A detailed comparison between the proposed MNN framework and a conventional fixed-threshold single-class Mahalanobis distance baseline is provided in Appendix~\ref{app:ch4_tables}. The baseline uses the 95th percentile of training-set Mahalanobis distances as a per-device acceptance threshold and has no knowledge of other device classes. As shown in the appendix, MNN achieves zero false acceptance for well-separated devices and drastically extends availability under noise.

\section{Adversarial Attacks and Empirical Thresholds}
\label{sec:adversarial}

This section evaluates the security of our authentication framework against white-box adversarial perturbations. The goal is to test whether the existing warning mechanism---based solely on $C_{\mathrm{claimed}}$ and the benign P5 threshold $\tau$---can detect gradient-based attacks, thereby revealing device-dependent empirical thresholds.

\subsection{Threat Model and Attack Design}
\label{sec:adv_threat}

\subsubsection{Adversarial attacks}
Gradient-based attacks such as the Fast Gradient Sign Method~\cite{Goodfellow2015} and Projected Gradient Descent (PGD)~\cite{Madry2018} craft small perturbations that flip classifier predictions. PGD is a standard and powerful white-box attack that iteratively maximizes the loss while projecting perturbations onto an $L_p$ ball; it is widely used to evaluate worst-case robustness. $L_2$ and $L_\infty$ constraints are the most common threat models. Sparse attacks modify only a limited subset of features~\cite{Papernot2016, Su2019}. 

\subsubsection{Adversary capabilities}
We assume a white-box adversary with full knowledge of the trained MNN classifier: the PCA projection matrix, centroids $\boldsymbol{\mu}_c$, regularised inverse covariances $\boldsymbol{\Sigma}_c^{-1}$, and per-feature standard deviations of the training set. The adversary intercepts a legitimate sample $\mathbf{x}$ from device $A$, adds a perturbation $\boldsymbol{\delta}$ to produce $\tilde{\mathbf{x}} = \mathbf{x} + \boldsymbol{\delta}$, and submits it with a claimed identity. The perturbation is further constrained to maintain physical plausibility.

\subsubsection{Perturbation constraints and the meaning of $\varepsilon$}
The perturbation magnitude is bounded by a parameter $\varepsilon$, which is scaled by the per-feature standard deviations of the training set. For $L_\infty$ attacks, each feature can be modified by at most $\pm\varepsilon$ times of its training-set standard deviation; for $L_2$ attacks, $\varepsilon$ is scaled by the global $L_2$ norm of the feature-wise standard deviation vector. In both cases, $\varepsilon$ represents the \emph{maximum allowed modification} relative to the natural variability of each feature: $\varepsilon = 1.0$ means the perturbation can reach the scale of typical measurement fluctuations, while $\varepsilon = 0.5$ corresponds to half of that scale. Unlike the random Gaussian noise in Section~\ref{sec:noise}, here the perturbation is adversarially optimised to fool the classifier.

\subsubsection{Attack types}
Five configurations are evaluated:
\begin{itemize}
    \item \textbf{$L_2$ targeted PGD:} the adversary minimises the Mahalanobis distance to a chosen target device $B \neq A$, under an $L_2$ norm constraint.
    \item \textbf{$L_\infty$ targeted PGD:} same objective, but the perturbation is constrained element-wise by an $L_\infty$ bound.
    \item \textbf{$L_2$ untargeted PGD:} the adversary maximises the distance to the true device $A$, aiming to cause any misclassification, under an $L_2$ constraint.
    \item \textbf{$L_\infty$ untargeted PGD:} same objective, under an $L_\infty$ constraint.
    \item \textbf{Sparse attack:} up to 200 features are modified greedily, one per iteration, according to the largest absolute gradient component. This simulates an adversary who can tamper with only a limited subset of measurement parameters.
\end{itemize}
All PGD attacks use 40 iterations and 2 random restarts with finite-difference gradient approximation. Targeted attacks randomly choose a target device different from the true device. Untargeted attacks are repeated 10 times per $\varepsilon$ for stable Warning Triggered Rate estimates.

\subsubsection{Defence mechanism}
No additional defence is added. The only protection is the warning mechanism of Section~\ref{sec:early_warning}: a sample claiming identity $c$ is accepted only if $C_{\mathrm{claimed}} \ge \tau_c$, rejected with alert if $0 < C_{\mathrm{claimed}} < \tau_c$, and rejected as misclassified if $C_{\mathrm{claimed}} \le 0$. Thresholds are identical to those used earlier: $\tau_{176}=0.292$, $\tau_{\mathrm{LQMS}}=0.729$, $\tau_{287}=0.171$.

\subsection{Experimental Setup and Metrics}
\label{sec:adv_setup}

The same chronological 70/30 train-test split is used. Perturbation strength $\varepsilon$ ranges from 0.1 to 3.0 (10 values). Evaluation metrics are:
\begin{itemize}
    \item \textbf{ASR (Attack Success Rate):} fraction classified as target device (targeted), or as any device different from the true device (untargeted).
    \item \textbf{Bypass Rate (targeted and sparse):} fraction classified as \emph{both} target \emph{and} satisfying $C_{\mathrm{claimed}} \ge \tau_{\mathrm{target}}$.
    \item \textbf{Warning Triggered Rate (untargeted only):} fraction with $C_{\mathrm{claimed}} < \tau_{\mathrm{true}}$ when evaluated against the true device. Combined with ASR, the Warning Triggered Rate provides a complete picture: a high rate on successful attacks indicates reliable detection of disruption, while a low rate on failed attacks confirms the absence of false alarms, meaning the system correctly remains silent when no classification error occurs.
\end{itemize}

\subsection{Results}
\label{sec:adv_results}

\subsubsection{$L_2$ Targeted Attacks}
Table~\ref{tab:adv_l2} reports representative results. ASR is high for all devices, confirming that $L_2$ gradient information effectively steers samples toward a target class. However, the Bypass Rate remains near zero across all perturbation levels: the vast majority of adversarial samples exhibit $C_{\mathrm{claimed}} < \tau_{\mathrm{target}}$, triggering the alert. The warning mechanism thus provides near-perfect detection against $L_2$ targeted attacks throughout the tested $\varepsilon$ range.

\begin{table}[!t]
\centering
\caption{$L_2$ targeted attack: ASR and Bypass Rate.}
\label{tab:adv_l2}
\footnotesize
\setlength{\tabcolsep}{5pt}
\begin{tabular}{l c c c c c c}
\toprule
\multirow{2}{*}{$\varepsilon$} & \multicolumn{2}{c}{\textbf{TianYan-176}} & \multicolumn{2}{c}{\textbf{LQMS-3}} & \multicolumn{2}{c}{\textbf{TianYan-287}} \\
\cmidrule(lr){2-3} \cmidrule(lr){4-5} \cmidrule(lr){6-7}
 & ASR & Bypass & ASR & Bypass & ASR & Bypass \\
\midrule
0.1 & 0.417 & 0.000& 0.000 & 0.000& 0.000 & 0.000\\
0.5 & 1.000 & 0.083& 0.484 & 0.000& 0.429 & 0.000\\
1.0 & 1.000 & 0.083& 0.355 & 0.000& 0.143 & 0.000\\
2.0 & 1.000 & 0.083& 0.516 & 0.000& 0.286 & 0.000\\
3.0 & 1.000 & 0.000& 0.613 & 0.000& 0.429 & 0.000\\
\bottomrule
\end{tabular}
\end{table}

\subsubsection{$L_\infty$ Targeted Attacks}
Table~\ref{tab:adv_linf} shows the results. At $\varepsilon = 0.1$ the attack is ineffective for all devices. As $\varepsilon$ grows, ASR rises, but the Bypass Rate remains well below unity for TianYan-176 and LQMS-3 through $\varepsilon = 0.7$, indicating that the warning still intercepts most successful attacks. For TianYan-287, ASR and Bypass Rate are both around 0.5 when $\varepsilon$ falls in $0.3$--$0.7$; the apparent non-monotonic fluctuation is a consequence of the small test set (7 samples, single-sample step $\approx 0.143$). Because $\tau_{287}=0.171$ is very low, once a sample is classified as TianYan-287, its $C_{\mathrm{claimed}}$ typically exceeds the threshold, so Bypass Rate naturally approaches ASR.

Experimentally, at $\varepsilon \ge 1.0$, ASR reaches 100\% for all devices and Bypass Rate rapidly approaches unity: the adversary can push samples deep into the target's core, producing high-confidence adversarial examples that evade the warning.

\begin{table}[!t]
\centering
\caption{$L_\infty$ targeted attack: ASR and Bypass Rate.}
\label{tab:adv_linf}
\footnotesize
\setlength{\tabcolsep}{5pt}
\begin{tabular}{l c c c c c c}
\toprule
\multirow{2}{*}{$\varepsilon$} & \multicolumn{2}{c}{\textbf{TianYan-176}} & \multicolumn{2}{c}{\textbf{LQMS-3}} & \multicolumn{2}{c}{\textbf{TianYan-287}} \\
\cmidrule(lr){2-3} \cmidrule(lr){4-5} \cmidrule(lr){6-7}
 & ASR & Bypass & ASR & Bypass & ASR & Bypass \\
\midrule
0.1 & 0.000 & 0.000 & 0.000 & 0.000 & 0.000 & 0.000 \\
0.3 & 0.000 & 0.000 & 0.000 & 0.000 & 0.571 & 0.429 \\
0.5 & 0.583 & 0.250 & 0.000 & 0.000 & 0.429 & 0.429 \\
0.7 & 0.750 & 0.333 & 0.000 & 0.000 & 0.571 & 0.571 \\
1.0 & 1.000 & 0.500 & 0.903 & 0.484 & 1.000 & 0.429 \\
1.5 & 1.000 & 1.000 & 1.000 & 0.903 & 1.000 & 1.000 \\
3.0 & 1.000 & 1.000 & 1.000 & 1.000 & 1.000 & 1.000 \\
\bottomrule
\end{tabular}
\end{table}

\subsubsection{Untargeted and Sparse Attacks}
$L_2$ untargeted attacks achieve 100\% ASR from $\varepsilon=0.2$ onward, with Warning Triggered Rate simultaneously 100\%---every successful denial-of-service is detected. $L_\infty$ untargeted attacks have 0\% ASR at all $\varepsilon$, and Warning Triggered Rate stays at the benign baseline (5--14\%, 14\% is due to the small sample size of TianYan-287), producing no excess false alarms. The sparse attack achieves 0\% ASR and 0\% Bypass Rate on all devices; modifying at most 200 of 1468 features is insufficient to override the global shape information preserved in the raw-curve fingerprint.

\subsection{Discussion}
\label{sec:adv_discussion}

The experiments reveal device-dependent empirical thresholds. The physical interpretation is grounded in the perturbation scale $\varepsilon$.

\subsubsection{Intact detection for $L_2$ targeted attacks}
Across the full tested range ($\varepsilon = 0.1$--$3.0$), the Bypass Rate for $L_2$ targeted attacks remains essentially zero. The gradient descent successfully pushes samples into the target's Mahalanobis Voronoi cell, but the resulting $C_{\mathrm{claimed}}$ consistently falls below $\tau_{\mathrm{target}}$, triggering the alert. No clear crossing point is observed within the tested range; the warning mechanism remains effective.

\subsubsection{Crossed threshold for $L_\infty$ targeted attacks}
The behaviour under $L_\infty$ targeted attacks is fundamentally different. At $\varepsilon = 0.5$, where the perturbation magnitude reaches half the natural feature standard deviation, the warning mechanism still intercepts the majority of attacks for TianYan-176 and all attacks for LQMS-3. Experimentally, at $\varepsilon \ge 1.0$, the perturbation exceeds the typical variability of the features, and the adversary can systematically produce high-confidence adversarial examples that bypass the warning.

The exact crossing point is device-dependent, reflecting the geometric margins established in Section~\ref{sec:confidence}: LQMS-3, with its large inter-class distances, resists bypass up to $\varepsilon = 1.0$; TianYan-287, with a low benign P5 of 0.171, exhibits non-negligible Bypass already at $\varepsilon = 0.3$. This device dependence is not a shortcoming of the framework but a direct consequence of the varying safety margins that the confidence distribution makes transparent.

\subsubsection{Physical feasibility}
$\varepsilon \ge 1.0$ represents a per-feature modification exceeding one standard deviation of natural fluctuations. In practice, such large, coordinated perturbations would likely be conspicuous under auxiliary physical monitoring of the raw measurement curves, providing an additional layer of defence beyond the confidence-based warning studied here.

\subsubsection{Devices with low confidence}
TianYan-287 exemplifies a device with inherently narrow geometric margin. Its low $\tau_{287}=0.171$ means that once an attack succeeds in classification, $C_{\mathrm{claimed}}$ easily clears the threshold. For such devices, stricter policies---higher percentile thresholds (e.g., P10 or P25), more frequent re-calibration, or combined physical anomaly detection---can raise the effective security boundary.

\subsubsection{Distinguishing Drift from Adversarial Attacks}
The current framework uses a static confidence threshold. Gradual natural drift and sudden adversarial perturbations both affect $C_{\mathrm{claimed}}$, but they exhibit different temporal signatures. Drift typically causes a slow change in the moving average of $C_{\mathrm{claimed}}$, whereas a successful adversarial attack introduces an abrupt, non-stationary deviation that can either increase or decrease $C_{\mathrm{claimed}}$ sharply relative to its recent history. Consequently, by monitoring a short-term moving average of $C_{\mathrm{claimed}}$ and the frequency of threshold violations, one can distinguish between benign drift and malicious attacks. This temporal discriminator is not implemented in the current work; its development and validation on multi-month hardware data are left as an important future direction.

\section{Conclusion and Future Work}
\label{sec:conclusion}

This work has presented a general framework for quantum hardware authentication based on multi-dimensional raw-curve fingerprints. By directly concatenating raw measurement statistics from four complementary experiments without curve fitting, we construct a 1468-dimensional fingerprint that preserves subtle device-specific information. A Mahalanobis nearest-neighbor classifier with Ledoit--Wolf shrinkage achieves 100\% benign accuracy on three superconducting processors over three weeks, and naturally yields an authentication confidence $C_{\mathrm{claimed}}$ that quantifies the geometric margin of each decision. The confidence score unifies benign diagnosis, drift monitoring, and adversarial defense: Its benign distribution reveals device-specific safety margins.  Under additive isotropic Gaussian noise it decays predictably to enable early warning. Against white-box attacks it detects $L_2$ targeted attacks with near-perfect success while revealing device-dependent empirical thresholds for $L_\infty$ attacks.

Several limitations and future directions should be noted: (i) the evaluation uses only three devices with modest sample sizes; validation on more devices and longer deployment periods is worth of studying. (ii) The fingerprint extraction protocol relies on a fixed experiment sequence; randomization or dummy measurements could mitigate potential adversarial identification. (iii) The $L_\infty$ attack can bypass the warning at $\varepsilon\ge1.0$; input-space monitoring or adversarial training may close this gap. (iv) Extending to open-set authentication (rejecting unknown devices) and developing a temporal discriminator to distinguish natural drift from adversarial attacks are important future directions.

\section{Acknowledgment}
\label{sec:acknowledgment}

This work was supported by the National Key Research and Development Program of China (Grant No. 2024YFB4504101).

\appendices

\section{Covariance Estimation}
\label{app:covariance}

For each device $c$, let $\mathbf{Z}_c \in \mathbb{R}^{n_c \times d}$ be the matrix of PCA-projected training samples ($n_c$ samples, $d=119$ components). The sample covariance matrix is $\mathbf{S}_c = \frac{1}{n_c-1} \mathbf{Z}_c^\top \mathbf{Z}_c$. The Ledoit--Wolf shrinkage estimator can be given by
\[
\hat{\boldsymbol{\Sigma}}_c = (1-\lambda_c)\mathbf{S}_c + \lambda_c \nu_c \mathbf{I},
\]
where the shrinkage intensity $\lambda_c \in [0,1]$ and the scaling factor $\nu_c$ are determined analytically from the data. To guarantee numerical invertibility, we add a small Tikhonov regularisation term $\gamma = 10^{-6}$ and use the regularised inverse
\[
\hat{\boldsymbol{\Sigma}}_c^{-1} = \bigl(\hat{\boldsymbol{\Sigma}}_c + \gamma \mathbf{I}\bigr)^{-1},
\]
calculated with the Moore--Penrose pseudoinverse if the matrix is ill‑conditioned.

\section{On the Role of Mahalanobis Distance in the Proposed Framework}
\label{app:mahalanobis}

The Mahalanobis distance is the central metric of our authentication framework. This appendix explains why it is chosen over alternative distances and how it enables the key properties reported in the paper.

\subsubsection{Definition and whitening interpretation}
For a point $\mathbf{z}$ (in the PCA subspace) and a class with centroid $\boldsymbol{\mu}_c$ and covariance $\boldsymbol{\Sigma}_c$, the squared Mahalanobis distance is $(\mathbf{z}-\boldsymbol{\mu}_c)^\top \boldsymbol{\Sigma}_c^{-1} (\mathbf{z}-\boldsymbol{\mu}_c)$. Factorising $\boldsymbol{\Sigma}_c^{-1} = \mathbf{L}^\top \mathbf{L}$ yields $\|\mathbf{L}(\mathbf{z}-\boldsymbol{\mu}_c)\|_2^2$, i.e., Euclidean distance after a linear whitening transformation that makes the class-conditional distribution isotropic (unit variance and zero correlations). Thus the Mahalanobis distance simultaneously normalises feature scales and removes linear dependencies.

\subsubsection{Why it is necessary for our fingerprint data}
Our 1468-dimensional raw fingerprint (Section~\ref{sec:feature_construction}) has three characteristics that make Euclidean or Manhattan distances unsuitable:
\begin{itemize}
    \item \textbf{Heterogeneous scales:} probabilities, gate counts, and time delays have vastly different natural variances. Without normalisation, high-variance dimensions would dominate the distance and mask subtle shape variations.
    \item \textbf{Strong correlations:} adjacent points on a Ramsey or SWAP curve are highly autocorrelated; $P_{01}$ and $P_{10}$ in SWAP are nearly complementary. Whitening automatically down-weights such redundant information.
    \item \textbf{Device-specific noise structures:} each quantum processor has its own variance profile (e.g., LQMS-3 has a larger spread along PC2 than TianYan-176). Using a \emph{class-conditional} covariance $\boldsymbol{\Sigma}_c$ allows each device to have its own distance metric, which is the essence of quadratic discriminant analysis (QDA).
\end{itemize}

\subsubsection{Relation to PCA and why down-weighting high within-class variance helps}
PCA is applied globally to the training set (all devices combined) and retains directions of large \emph{total} variance (95\% cumulative). A direction may have large total variance either because it captures genuine differences between devices (useful for classification) or because it contains common noise that affects all devices (not useful). Importantly, a direction with large total variance can still have large \emph{within-class} variance for a particular device---meaning repeated measurements of that device fluctuate widely along that direction. Such directions are unreliable for authentication. The Mahalanobis distance, using a per-device covariance estimate, scales each direction by the inverse of its within-class standard deviation. Consequently, directions with large within-class variance are \emph{down-weighted}, while directions with small within-class variance (stable, reproducible features) are relatively amplified. In our data, the inter-device differences predominantly lie in low-within-variance directions (stable physical fingerprints), and high-within-variance directions correspond mainly to measurement noise or environmental fluctuations. Therefore, down-weighting the latter improves separability. This is complementary to PCA: PCA discards globally negligible directions, and Mahalanobis suppresses within-class noise among the retained ones.

\subsubsection{Specific roles in the framework}
\begin{itemize}
    \item \textbf{Authentication:} The MNN classifier assigns a test sample to the device with smallest Mahalanobis distance. The class-conditional whitening yields clean geometric separation (inter-/intra-class ratios $>3$), giving 100\% accuracy under natural drift.
    \item \textbf{Confidence score $C_{\mathrm{claimed}}$:} Defined as $1 - D_{\mathrm{claimed}}/D_{\mathrm{second}}$, the ratio of Mahalanobis distances is meaningful only after whitening, because distances from different classes are then expressed in units of their own variability. This provides a device-specific ``safety margin''.
    \item \textbf{Predictable noise decay:} Under additive isotropic Gaussian noise, the expected Mahalanobis distance grows as $\sqrt{D^2(0) + \sigma^2 \operatorname{tr}(\boldsymbol{\Sigma}_c^{-1})}$ (Appendix~\ref{app:theory}). The trace term $\operatorname{tr}(\boldsymbol{\Sigma}_c^{-1})$ quantifies device sensitivity, explaining why TianYan-176 decays fast while TianYan-287 barely decays. This enables the early-warning mechanism.
    \item \textbf{Adversarial detection:} The same confidence threshold detects $L_2$ targeted attacks with near-zero bypass rate (Section~\ref{sec:adversarial}) and reveals device-dependent $L_\infty$ thresholds. The whitened geometry makes it hard to produce high-confidence adversarial examples without exceeding physically plausible perturbation sizes.
\end{itemize}

\section{Theoretical Analysis of $C_{\mathrm{claimed}}$ under Isotropic Gaussian Noise}
\label{app:theory}

\subsection{Expected Mahalanobis Distance and Confidence}
\label{app:expectation}

Let the PCA subspace have dimension $d$  (here $d=119$). For a device $c$ we have the training centroid $\boldsymbol{\mu}_c$ and the regularised inverse covariance matrix $\boldsymbol{\Sigma}_c^{-1}$ (positive definite). Consider a test sample $\mathbf{z}$ that truly belongs to device $c$. Its squared Mahalanobis distance to its own centroid and to another centroid $c'$ are
\begin{gather*}
D_{c}^{2}(0) = (\mathbf{z}-\boldsymbol{\mu}_c)^\top \boldsymbol{\Sigma}_c^{-1} (\mathbf{z}-\boldsymbol{\mu}_c), \\
D_{c'}^{2}(0) = (\mathbf{z}-\boldsymbol{\mu}_{c'})^\top \boldsymbol{\Sigma}_{c'}^{-1} (\mathbf{z}-\boldsymbol{\mu}_{c'}).
\end{gather*}
Add isotropic Gaussian noise $\boldsymbol{\eta} \sim \mathcal{N}(\mathbf{0}, \sigma^2 \mathbf{I}_d)$, independent of $\mathbf{z}$. Then
\[
\tilde{\mathbf{z}} = \mathbf{z} + \boldsymbol{\eta},
\]
and
\[
R_{c'}^2 = \|\tilde{\mathbf{z}}-\boldsymbol{\mu}_{c'}\|_{\boldsymbol{\Sigma}_{c'}^{-1}}^2 = D_{c'}^2(0) + 2\boldsymbol{\eta}^\top \boldsymbol{\Sigma}_{c'}^{-1}(\mathbf{z}-\boldsymbol{\mu}_{c'}) + \boldsymbol{\eta}^\top \boldsymbol{\Sigma}_{c'}^{-1}\boldsymbol{\eta}.
\]
Taking expectation conditional on $\mathbf{z}$, the cross term vanishes because $\mathbb{E}[\boldsymbol{\eta}]=\mathbf{0}$. Using $\mathbb{E}[\boldsymbol{\eta}^\top \mathbf{A}\boldsymbol{\eta}] = \sigma^2 \operatorname{tr}(\mathbf{A})$ for any deterministic symmetric matrix $\mathbf{A}$, we obtain the exact expression
\[
\mathbb{E}[R_{c'}^2 \mid \mathbf{z}] = D_{c'}^2(0) + \sigma^2 \operatorname{tr}(\boldsymbol{\Sigma}_{c'}^{-1}), \qquad \forall c'. 
\]

The expectation of the square root is not exactly the square root of the expectation. A first-order Delta method expansion gives
\begin{align*}
\mathbb{E}\bigl[\sqrt{R_{c'}^2}\bigm|\mathbf{z}\bigr] 
&= \sqrt{\mathbb{E}[R_{c'}^2\mid\mathbf{z}]} \;-\; \frac{\operatorname{Var}(R_{c'}^2\mid\mathbf{z})}{8\,\bigl(\mathbb{E}[R_{c'}^2\mid\mathbf{z}]\bigr)^{3/2}} + \cdots .
\end{align*}
In the high-dimensional regime considered here ($d=119$), the squared coefficient of variation of $R_{c'}^2$ is small due to the concentration of quadratic forms of Gaussian vectors, making the correction term negligible compared to the leading square-root term. We therefore adopt the approximation
\[
\widehat{D}_{c'}(\sigma) := \sqrt{ \mathbb{E}[R_{c'}^2 \mid \mathbf{z}] } = \sqrt{ D_{c'}^2(0) + \sigma^2 \operatorname{tr}(\boldsymbol{\Sigma}_{c'}^{-1}) }.
\]
The predicted authentication confidence for the given sample, assuming that it truthfully claims identity $c$, is then
\[
\widehat{C}_{\mathrm{claimed}}(\sigma) = 1 - \frac{ \widehat{D}_{c}(\sigma) }{ \displaystyle\min_{c'\neq c} \widehat{D}_{c'}(\sigma) }.
\]

Averaging $\widehat{C}_{\mathrm{claimed}}(\sigma)$ over all test samples of device $c$ yields the theoretical curve that is compared with the experimental mean in Section~\ref{sec:iso_results}. The agreement is close despite the omission of the variance term, confirming that the approximation captures the dominant noise effect. The formula involves only the noise-free distances of the test samples and the training-set traces $\operatorname{tr}(\boldsymbol{\Sigma}_{c'}^{-1})$, which are directly computed from the training data. This derivation does not assume any particular distribution of the test samples other than the additive Gaussian noise, and it naturally accounts for the natural drift already present in the test samples through the measured $D_{c'}(0)$. Note that if $\sigma$ becomes large enough that the nearest competing centroid changes (i.e., $\min_{c'\neq c} \widehat{D}_{c'}(\sigma)$ is attained for a different $c'$), the expression above automatically handles the switch because the minimum is recomputed at each $\sigma$.

\subsection{Monotonicity Condition for Individual Samples}
\label{app:monotonicity}

For a fixed claimed device $c$ and a given test sample, let
\begin{align*}
a &= D_{c}^{2}(0), \quad b = \operatorname{tr}(\boldsymbol{\Sigma}_{c}^{-1}), \\
c &= D_{\mathrm{second}}^{2}(0), \quad d = \operatorname{tr}(\boldsymbol{\Sigma}_{\mathrm{second}}^{-1}),
\end{align*}
where ``second'' denotes the nearest competing device under noise-free conditions (the competitor may change with a large $\sigma$, but for the monotonicity analysis we consider the initial competitor that determines the behaviour with a small to moderate $\sigma$). The squared ratio of expected distances is
\[
r^{2}(\sigma) = \frac{a + b\sigma^{2}}{c + d\sigma^{2}}.
\]
Differentiating with respect to $\sigma^{2}$ gives
\[
\frac{d(r^{2})}{d(\sigma^{2})} = \frac{bc - ad}{(c + d\sigma^{2})^{2}}.
\]
Since the denominator is always positive, the sign of the derivative is determined by
\[
\Delta = bc - ad.
\]
Hence:
\begin{itemize}
    \item If $\Delta > 0$, the ratio $r(\sigma)$ increases monotonically, and $C_{\mathrm{claimed}}$ decreases monotonically.
    \item If $\Delta < 0$, the ratio decreases monotonically, and $C_{\mathrm{claimed}}$ increases monotonically.
    \item If $\Delta = 0$, the ratio is constant.
\end{itemize}

\subsubsection{Application to the three devices}
Using the per-device traces and typical noise-free distances (Section~\ref{sec:iso_results}), we obtain the following picture.

\begin{itemize}
    \item \textbf{TianYan-176} ($b=52.11$, $d=26.68$, typical $a\approx49.7$, $c\approx419.0$): 
    \[
    \Delta = 52.11\times419.0 - 49.7\times26.68 \gg 0.
    \]
    For all test samples $\Delta$ is strongly positive; therefore every individual sample exhibits a monotonic decrease of $C_{\mathrm{claimed}}$, producing the fast drop observed in Fig.~\ref{fig:c_vs_sigma}.
    
    \item \textbf{LQMS-3} ($b=36.88$, $d=26.68$, typical $a\approx49.8$, $c\approx1490$): 
    \[
    \Delta > 0 \text{ for all samples},
    \]
    but the large initial competitor distance $c$ makes the ratio change very slowly, resulting in the gradual, strictly decreasing curve.
    
    \item \textbf{TianYan-287} ($b=26.68$, $d=52.11$, $a$ and $c$ vary widely across test samples). Here 
    \[
    \Delta = 26.68\,c - 52.11\,a.
    \]
    The critical condition $\Delta = 0$ yields $a \approx (26.68/52.11)\,c$. With a typical competitor distance $c\approx419.0$, the critical initial self-distance is $D_{c}(0) \approx 14.65$. Samples with $D_{c}(0) < 14.65$ have $\Delta > 0$ and show decreasing confidence; samples with $D_{c}(0) > 14.65$ have $\Delta < 0$ and show slowly increasing confidence. The test set of TianYan-287 contains a mixture of both types, leading to an average curve that is almost flat, with a slight non-monotonic fluctuation well within one standard deviation. This behaviour is fully consistent with the experimental data and does not compromise the overall predictive power of the threshold-based warning.
\end{itemize}

Thus, the trace-based analysis not only explains the average decay rates but also provides a precise condition for individual monotonicity, giving a complete theoretical account of the observed $C_{\mathrm{claimed}}$ curves under isotropic Gaussian noise.

\section{Additional Material for Section~IV: Comparison with a Single-Class Baseline}
\label{app:ch4_tables}

This appendix provides the complete experimental details and results of the comparison between the proposed MNN framework and a conventional fixed-threshold single-class Mahalanobis distance baseline, which was described in Section~\ref{sec:comp_baseline}.

\subsection{Motivation and baseline construction}
The comparison serves to quantify the benefit of multi-class information in our authentication framework. The single-class baseline represents the most natural way to perform authentication using Mahalanobis distance when only one device's training data is available. For each device $c$, the baseline is constructed as follows:
\begin{enumerate}
    \item Compute the Mahalanobis distance from every training sample of device $c$ to its own centroid $\boldsymbol{\mu}_c$.
    \item Take the 95th percentile of these distances as the acceptance threshold $\theta_c$.
    \item During testing, a sample claiming identity $c$ is accepted if its Mahalanobis distance to $\boldsymbol{\mu}_c$ is $\le\theta_c$, and rejected otherwise.
\end{enumerate}
This model has no knowledge of other device classes; it makes an independent accept/reject decision for each claimed identity without a joint classification rule. Consequently, it does not produce a single overall multi-class accuracy.

\subsection{Evaluation metrics}
Both the baseline and the MNN classifier are evaluated under the same additive isotropic Gaussian noise with 10 independent realisations per noise strength $\sigma$. We report:
\begin{itemize}
    \item \textbf{FAR (False Acceptance Rate):} the fraction of samples not belonging to device $c$ that are incorrectly accepted as $c$.
    \item \textbf{FRR (False Rejection Rate):} the fraction of genuine samples of device $c$ that are incorrectly rejected.
    \item \textbf{Overall Accuracy (MNN only):} the fraction of all test samples correctly classified, reported only for MNN since the single-class baseline makes independent per-device decisions without a joint prediction.
\end{itemize}

\subsection{Complete results}
Table~\ref{tab:comparison_full} reports the full comparison across all noise strengths $\sigma = 0,1,\dots,10$. Values are mean $\pm$ standard deviation over 10 noise realisations.

\begin{table*}[!t]
\centering
\caption{Complete comparison between MNN and the fixed-threshold single-class baseline. Values are mean $\pm$ std over 10 noise realisations.}
\label{tab:comparison_full}
\scriptsize
\setlength{\tabcolsep}{2pt}
\renewcommand{\arraystretch}{0.9}
\begin{tabular}{l c c c c c c}
\toprule
\multirow{2}{*}{Device} & \multirow{2}{*}{$\sigma$} & \multicolumn{2}{c}{\textbf{Single-class}} & \multicolumn{3}{c}{\textbf{MNN}} \\
\cmidrule(lr){3-4} \cmidrule(lr){5-7}
 & & FAR & FRR & Overall Acc. & FAR & FRR \\
\midrule
\multirow{10}{*}{TianYan-176}
& 0 & $0.158\pm0.000$ & $0.000\pm0.000$ & $1.000\pm0.000$ & $0.000\pm0.000$ & $0.000\pm0.000$ \\
& 1 & $0.158\pm0.000$ & $0.000\pm0.000$ & $1.000\pm0.000$ & $0.000\pm0.000$ & $0.000\pm0.000$ \\
& 2 & $0.158\pm0.000$ & $0.000\pm0.000$ & $1.000\pm0.000$ & $0.000\pm0.000$ & $0.000\pm0.000$ \\
& 3 & $0.141\pm0.017$ & $0.000\pm0.000$ & $0.962\pm0.020$ & $0.000\pm0.000$ & $0.158\pm0.083$ \\
& 4 & $0.088\pm0.031$ & $0.013\pm0.030$ & $0.841\pm0.033$ & $0.000\pm0.000$ & $0.663\pm0.138$ \\
& 5 & $0.045\pm0.029$ & $0.225\pm0.109$ & $0.777\pm0.017$ & $0.000\pm0.000$ & $0.929\pm0.071$ \\
& 6 & $0.018\pm0.019$ & $0.571\pm0.152$ & $0.763\pm0.007$ & $0.000\pm0.000$ & $0.988\pm0.030$ \\
& 7 & $0.004\pm0.009$ & $0.863\pm0.122$ & $0.760\pm0.000$ & $0.000\pm0.000$ & $1.000\pm0.000$ \\
& 8 & $0.001\pm0.006$ & $0.971\pm0.040$ & $0.760\pm0.000$ & $0.000\pm0.000$ & $1.000\pm0.000$ \\
& 10& $0.000\pm0.000$ & $1.000\pm0.000$ & $0.760\pm0.000$ & $0.000\pm0.000$ & $1.000\pm0.000$ \\
\midrule
\multirow{10}{*}{LQMS-3}
& 0 & $0.000\pm0.000$ & $0.000\pm0.000$ & $1.000\pm0.000$ & $0.000\pm0.000$ & $0.000\pm0.000$ \\
& 1 & $0.000\pm0.000$ & $0.000\pm0.000$ & $1.000\pm0.000$ & $0.000\pm0.000$ & $0.000\pm0.000$ \\
& 2 & $0.000\pm0.000$ & $0.018\pm0.024$ & $1.000\pm0.000$ & $0.000\pm0.000$ & $0.000\pm0.000$ \\
& 3 & $0.000\pm0.000$ & $0.660\pm0.079$ & $1.000\pm0.000$ & $0.000\pm0.000$ & $0.000\pm0.000$ \\
& 4 & $0.000\pm0.000$ & $1.000\pm0.000$ & $1.000\pm0.000$ & $0.000\pm0.000$ & $0.000\pm0.000$ \\
& 5 & $0.000\pm0.000$ & $1.000\pm0.000$ & $1.000\pm0.000$ & $0.000\pm0.000$ & $0.000\pm0.000$ \\
& 6 & $0.000\pm0.000$ & $1.000\pm0.000$ & $1.000\pm0.000$ & $0.000\pm0.000$ & $0.000\pm0.000$ \\
& 7 & $0.000\pm0.000$ & $1.000\pm0.000$ & $0.999\pm0.004$ & $0.000\pm0.000$ & $0.002\pm0.007$ \\
& 8 & $0.000\pm0.000$ & $1.000\pm0.000$ & $0.996\pm0.010$ & $0.000\pm0.000$ & $0.007\pm0.016$ \\
& 10& $0.000\pm0.000$ & $1.000\pm0.000$ & $0.925\pm0.037$ & $0.000\pm0.000$ & $0.121\pm0.059$ \\
\midrule
\multirow{10}{*}{TianYan-287}
& 0 & $0.000\pm0.000$ & $0.571\pm0.000$ & $1.000\pm0.000$ & $0.000\pm0.000$ & $0.000\pm0.000$ \\
& 1 & $0.000\pm0.000$ & $0.571\pm0.000$ & $1.000\pm0.000$ & $0.000\pm0.000$ & $0.000\pm0.000$ \\
& 2 & $0.000\pm0.000$ & $0.529\pm0.080$ & $0.947\pm0.031$ & $0.062\pm0.036$ & $0.000\pm0.000$ \\
& 3 & $0.000\pm0.000$ & $0.543\pm0.057$ & $0.783\pm0.024$ & $0.252\pm0.028$ & $0.000\pm0.000$ \\
& 4 & $0.000\pm0.000$ & $0.586\pm0.089$ & $0.761\pm0.004$ & $0.278\pm0.005$ & $0.000\pm0.000$ \\
& 5 & $0.000\pm0.000$ & $0.557\pm0.089$ & $0.714\pm0.027$ & $0.333\pm0.031$ & $0.000\pm0.000$ \\
& 6 & $0.000\pm0.000$ & $0.671\pm0.102$ & $0.546\pm0.053$ & $0.528\pm0.062$ & $0.000\pm0.000$ \\
& 7 & $0.000\pm0.000$ & $0.771\pm0.123$ & $0.339\pm0.051$ & $0.769\pm0.060$ & $0.000\pm0.000$ \\
& 8 & $0.000\pm0.000$ & $0.821\pm0.135$ & $0.209\pm0.027$ & $0.920\pm0.032$ & $0.000\pm0.000$ \\
& 10& $0.000\pm0.000$ & $0.907\pm0.093$ & $0.145\pm0.011$ & $0.994\pm0.013$ & $0.000\pm0.000$ \\
\bottomrule
\end{tabular}
\end{table*}

\subsection{Key findings}
The comparison reveals three main points that support the advantage of multi-class information:
\begin{itemize}
    \item \textbf{Security.} MNN maintains zero FAR for TianYan-176 and LQMS-3 at all noise levels, whereas the single-class baseline already shows 15.8\% FAR for TianYan-176 even without noise. The multi-class constraint inherently prevents impostor acceptance.
    \item \textbf{Availability.} For LQMS-3, the single-class baseline rejects nearly all genuine samples from $\sigma=4$ onward, while MNN retains 100\% accuracy up to $\sigma=6$ and still reaches 92.5\% at $\sigma=10$. The multi-class decision boundary dramatically extends the usable lifetime.
    \item \textbf{Compact attractor effect.} MNN's FAR for TianYan-287 increases with noise because noisy samples from other devices drift into its compact region. This is a geometric consequence, not a vulnerability of TianYan-287 itself (its FRR remains zero). The single-class baseline is unusable for this device even in benign conditions (FRR 57.1\% due to the small training set).
\end{itemize}

\subsection{Early warning state proportions}
Table~\ref{tab:early_warning_full} reports the complete proportion of samples in Safe, Warning, and Error states for all noise strengths, complementing the condensed version in Section~\ref{sec:early_warning}.

\begin{table}[!htbp]
\centering
\scriptsize
\setlength{\tabcolsep}{3pt}
\renewcommand{\arraystretch}{0.9}
\begin{threeparttable}
\caption{Proportion of samples in Safe, Warning, and Error states for all devices and $\sigma$ values. Threshold $\tau_c$ = benign P5.}
\label{tab:early_warning_full}
\begin{tabular}{l c c c c c c c c c c c c}
\toprule
\multirow{2}{*}{\textbf{Device ($\tau_c$)}} & \multirow{2}{*}{\textbf{State}} & \multicolumn{11}{c}{\textbf{Noise level $\sigma$}} \\
\cmidrule(lr){3-13}
 & & 0 & 1 & 2 & 3 & 4 & 5 & 6 & 7 & 8 & 9 & 10 \\
\midrule
\multirow{3}{*}{TianYan-176 (0.292)}
& Safe & 0.917 & 0.558 & 0.000 & 0.000 & 0.000 & 0.000 & 0.000 & 0.000 & 0.000 & 0.000 & 0.000 \\
& Warning & 0.083 & 0.442 & 0.742 & 0.008 & 0.000 & 0.000 & 0.000 & 0.000 & 0.000 & 0.000 & 0.000 \\
& Error & 0.000 & 0.000 & 0.258 & 0.992 & 1.000 & 1.000 & 1.000 & 1.000 & 1.000 & 1.000 & 1.000 \\
\midrule
\multirow{3}{*}{LQMS-3 (0.729)}
& Safe & 0.936 & 0.132 & 0.000 & 0.000 & 0.000 & 0.000 & 0.000 & 0.000 & 0.000 & 0.000 & 0.000 \\
& Warning & 0.064 & 0.868 & 1.000 & 1.000 & 1.000 & 1.000 & 0.984 & 0.877 & 0.774 & 0.623 & 0.423 \\
& Error & 0.000 & 0.000 & 0.000 & 0.000 & 0.000 & 0.000 & 0.016 & 0.123 & 0.226 & 0.377 & 0.577 \\
\midrule
\multirow{3}{*}{TianYan-287 (0.171)}
& Safe & 0.857 & 0.786 & 0.829 & 0.886 & 0.971 & 0.971 & 0.986 & 0.986 & 0.957 & 0.929 & 0.900 \\
& Warning & 0.143 & 0.214 & 0.171 & 0.114 & 0.029 & 0.029 & 0.014 & 0.014 & 0.043 & 0.071 & 0.100 \\
& Error & 0.000 & 0.000 & 0.000 & 0.000 & 0.000 & 0.000 & 0.000 & 0.000 & 0.000 & 0.000 & 0.000 \\
\bottomrule
\end{tabular}
\end{threeparttable}
\end{table}

\end{document}